\documentclass[journal]{IEEEtran}

\usepackage{amsmath,amsfonts}
\usepackage[justification=justified,font=small]{caption}	
\usepackage[labelformat=simple]{subcaption}

\usepackage{textcomp}
\usepackage{stfloats}
\usepackage{url}
\usepackage{verbatim}
\usepackage{graphicx}
\usepackage{cite}
\usepackage{setspace}
\usepackage{epsfig}  
\usepackage{graphicx}  
\usepackage{amssymb}  
\usepackage{lscape}  
\usepackage{color}
\usepackage{mathrsfs}
\usepackage{mathtools}
\usepackage{amsthm}
\usepackage{array}
\usepackage{booktabs}

\usepackage{longtable}

\usepackage{cases}
\usepackage{bm}
\usepackage{textcomp}
\newcommand{\textapprox}{\raisebox{0.5ex}{\texttildelow}}

\graphicspath{{./figures/}}

\begin{document}
\title{Analytical Quantum Full-Wave Solutions for a 3D Circuit Quantum Electrodynamics System}
%
%
%

\author{Soomin Moon,~\IEEEmembership{Graduate Student Member,~IEEE,}
        Dong-Yeop~Na,~\IEEEmembership{Member,~IEEE,}
        and~Thomas~E.~Roth,~\IEEEmembership{Member,~IEEE}
\thanks{Manuscript received XXXX XX, 2024. \\
\indent This work is based upon work supported by the National Science Foundation under Grant No. 2202389 and Purdue ECE Elmore Emerging Frontiers Center ``The Crossroads of Quantum and AI.''. \textit{(Corresponding author: Thomas E. Roth.)}\\
\indent S. Moon and T. E. Roth are with the Elmore Family School of Electrical and Computer Engineering, Purdue University, West Lafayette, IN 47907 USA and the Purdue Quantum Science and Engineering Institute, West Lafayette, IN 47907 USA (e-mail: rothte@purdue.edu).\\
\indent D.-Y. Na is with the Department of Electrical and Computer Engineering, Pohang University of Science and Technology (POSTECH), Pohang 37673, South Korea (e-mail: dyna22@postech.ac.kr).\\
\indent This work has been submitted to the IEEE for possible publication. Copyright may be transferred without notice, after which this version may no longer be accessible.} 
}

%
%

\markboth{Journal of \LaTeX\ Class Files,~Vol.~X, No.~X, XXXX~2024}%
{Shell \MakeLowercase{\textit{et al.}}: Bare Demo of IEEEtran.cls for IEEE Journals}
%



\maketitle

\begin{abstract}
High-fidelity general-purpose numerical methods are increasingly needed to improve superconducting circuit quantum information processor performance. One challenge in developing such numerical methods is the lack of reference data to validate them. To address this, we have designed a 3D system where all electromagnetic properties needed in a quantum analysis can be evaluated using analytical techniques from classical electromagnetic theory. Here, we review the basics of our field-based quantization method and then use these techniques to create the first-ever analytical quantum full-wave solution for a superconducting circuit quantum device. Specifically, we analyze a coaxial-fed 3D waveguide cavity with and without transmon quantum bits inside the cavity. We validate our analytical solutions by comparing them to numerical methods in evaluating single photon interference and computing key system parameters related to controlling quantum bits. In the future, our analytical solutions can be used to validate numerical methods, as well as build intuition about important quantum effects in realistic 3D devices.
\end{abstract}

\begin{IEEEkeywords}
Circuit quantum electrodynamics, transmon qubit, quantum theory, microwave resonators, cavity perturbation theory, and antenna theory.
\end{IEEEkeywords}

%
\IEEEpeerreviewmaketitle


%
\section{Introduction}
\label{sec:introduction}
\IEEEPARstart{T}{he} superconducting circuit platform is a leading approach for developing quantum computers \cite{arute2019quantum,wu2021strong} and other quantum information processing technologies \cite{gu2017microwave,krantz2019quantum,blais2021circuit} that are expected to revolutionize many areas of science and technology. Typically referred to as circuit quantum electrodynamics (cQED) devices, these systems utilize the interactions between microwave electromagnetic (EM) fields and superconducting circuits to generate and process quantum information. Despite significant experimental progress, substantial improvements are still needed for these technologies to be useful in practice. For instance, quantum computers have achieved a quantum advantage over classical supercomputers \cite{arute2019quantum,wu2021strong}, but for such an advantage to be realized on realistic problems requires improving the performance of most components while massively scaling the number of quantum bits (qubits) in the system \cite{jurcevic2021demonstration,acharya2023suppressing}. To overcome these engineering challenges, general-purpose and high-fidelity numerical analysis tools are becoming increasingly important \cite{nigg2012black,solgun2014blackbox,minev2021energy,roth2021macroscopic}.

Unfortunately, there are currently only a few general-purpose numerical methods available for modeling cQED devices and they suffer from significant inefficiencies. Some of the first general-purpose numerical approaches were blackbox circuit quantization methods \cite{nigg2012black,solgun2014blackbox}, which use full-wave EM simulations of all linear components to compute the impedance matrix that is then used to build an equivalent Foster circuit network. This equivalent circuit is then quantized in conjunction with the remaining nonlinear aspects of the qubits to characterize the full system. In practice, these methods often require an inconvenient trade-off between accuracy and user-intensive curve-fitting procedures that can require performing refined simulations around any resonant peaks in the multi-port impedance matrix. More recently, the energy participation ratio (EPR) quantization method \cite{minev2021energy} was introduced as an alternative to blackbox circuit quantization. This approach recasts the theoretical description of the system so that instead of performing impedance matrix simulations the results of full-wave EM eigenmode decompositions of the linear part of a cQED device are used to quantize the system. As a result, this method avoids user-intensive curve-fitting procedures. 

Regardless of whether blackbox circuit or EPR quantization is used, the manner in which the linear and nonlinear parts of the qubits in the cQED system are subdivided requires the use of many quantum states per resonant mode to reach numerical convergence, as will be shown later. This severely limits the size of cQED system that can be analyzed because the dimension of the matrix needed to characterize the quantum aspects of the system grows exponentially with respect to the number of quantum states per resonant mode.

An alternative method to these quantization approaches is our macroscopic cQED formalism proposed in \cite{roth2021macroscopic} that utilizes a field theory description of a system. In this approach, key parameters in the quantum description of a system can also be evaluated in terms of the results of a linear EM eigenmode analysis. However, due to how the nonlinearity of qubits are incorporated into this formalism, a much smaller number of quantum states per mode can be used to reach numerical convergence for an accurate solution. 

However, for both EPR and our field-based quantization methods, EM eigenmodes must be found numerically which becomes computationally prohibitive for large devices. As a result, there is a need for more efficient methods to be formulated. Unfortunately, validating new numerical methods in this field is a challenge itself due to the lack of reference data. In the case of measured data, the manufacturing precision and presence of other uncontrolled factors in experiments limits the achievable level of quantitative validation \cite{minev2021energy}. Further, access to measured data is much more limited than for traditional EM applications. Another main avenue for validation in typical computational electromagnetics applications is the use of analytical solutions, such as the Mie series for spherical scattering. However, a similar analytical solution is not currently available for cQED systems. To help address this, we have designed a simple geometry for which all field-based aspects of the quantum Hamiltonian proposed in \cite{roth2021macroscopic} can be evaluated analytically using results from cavity perturbation theory \cite{pozar2009microwave} and antenna theory \cite{balanis2016antenna}. In the future, these analytical solutions can be used to validate new numerical methods, as well as build intuition about important quantum effects that occur in realistic cQED devices. 

More specifically, we consider a system inspired by \textit{3D transmons} \cite{paik2011observation}. These consist of a transmon qubit \cite{koch2007charge,roth2022transmon} formed by a small planar dipole antenna that is embedded in a waveguide cavity. We consider a similar system with and without transmon qubits inside a coaxially-fed rectangular waveguide cavity to develop analytical solutions. For the case without transmon qubits, we construct our analytical solution in the context of quantum input-output theory \cite{walls2007quantum} to consider how different quantum input states are scattered through the system. We use this solution to analyze the Hong-Ou-Mandel (HOM) effect \cite{hong1987measurement,na2020classical} between single photons scattering through the cavity system, which is often used to qualitatively validate computational quantum electromagnetics methods \cite{na2020quantum,na2021diagonalization,na2022numerical}. For the case with transmon qubits present, we calculate key system parameters that are important for the control and measurement of qubit states. We validate our analytical solutions by comparing our results to our formalism using numerical EM eigenmodes, to EPR quantization, or other impedance-based analysis methods \cite{solgun2019simple,solgun2022direct}.  

Preliminary results on these analytical solutions were reported in \cite{moon2023validation,moon2023full,moon2024analysis}. This work expands on \cite{moon2023validation,moon2023full} by generalizing the analysis to a fully-quantum setting. We provide additional details on the quantum input-output theory method, discuss how to compute the relevant parameters analytically for our geometry, and present how to use the quantum input-output theory results to compute Hong-Ou-Mandel interference curves. This work expands on \cite{moon2024analysis} by providing more details on the derivation of the analytical solution, considers the multi-qubit case, and presents new numerical results.   

The remainder of this work is organized in the following manner. In Section \ref{sec:background}, we review our macroscopic cQED approach in the context of the system geometries analyzed here. Then, in Section \ref{sec:empty-cavity-analytical} we discuss how to use quantum input-output theory and traditional EM methods to analytically model HOM interference effects in a port-fed cavity. Next, we discuss in Section \ref{sec:qubit-analytical} how to use traditional EM methods to evaluate all the field-based parameters in our Hamiltonian characterizing transmon qubits placed inside a closed cavity. Results are presented at the ends of Sections \ref{sec:empty-cavity-analytical} and \ref{sec:qubit-analytical} to validate the respective analytical solutions. Finally, we discuss conclusions and future work in Section \ref{sec:conclusion}.
\section{Macroscopic Circuit Quantum Electrodynamics}
\label{sec:background}
Before presenting our analytical solutions, it is necessary to discuss how to apply the macroscopic cQED formalism of \cite{roth2021macroscopic} to the systems considered in this work. In this approach, field quantization is performed in the framework of macroscopic quantum electrodynamics \cite{scheel2008macroscopic} where lossless, non-dispersive media are considered in terms of macroscopic quantities like permittivity rather than through microscopic descriptions (introductions to field quantization in the macroscopic context can be found in \cite{chew2016quantum,chew2016quantum2,chew2021qme-made-simple}). Further, as is common in cQED systems, \cite{roth2021macroscopic} treats properties of the qubits in terms of macroscopic degrees of freedom rather than utilizing a full microscopic description of the superconducting materials. 

In this section, we discuss two cases relevant to the analytical solutions developed here. In Section \ref{subsec:background-w-ports}, we consider the quantum description of an empty cavity fed by waveguiding ports. In Section \ref{subsec:background-cavity+qubit}, we consider a transmon qubit located inside a closed cavity. These cases can be combined for a more complete description of a general system \cite{roth2021macroscopic}, but this then requires a numerical solution to analyze specific dynamical scenarios, which is outside of the scope of this work. 

\subsection{Port-Fed Empty Cavity}
\label{subsec:background-w-ports}
\begin{figure}[t]
    \centering
    \begin{subfigure}[t]{0.9\linewidth}
        \includegraphics[width=\textwidth]{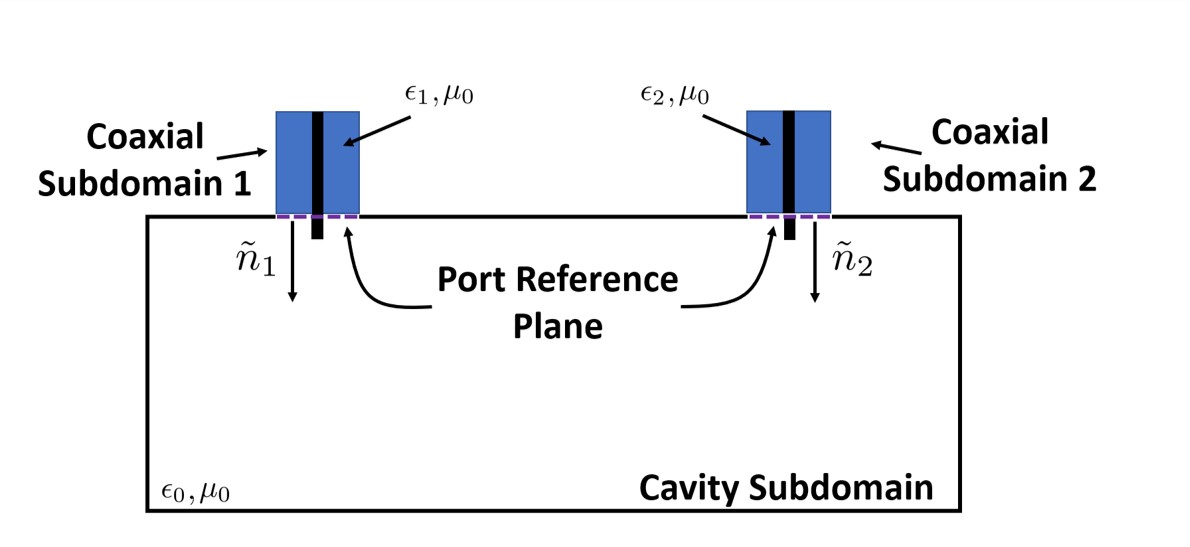}
        \caption{}
        \label{subfig:geometry_empty_system}
    \end{subfigure}
    \begin{subfigure}[t]{0.9\linewidth}
		\includegraphics[width=\textwidth]{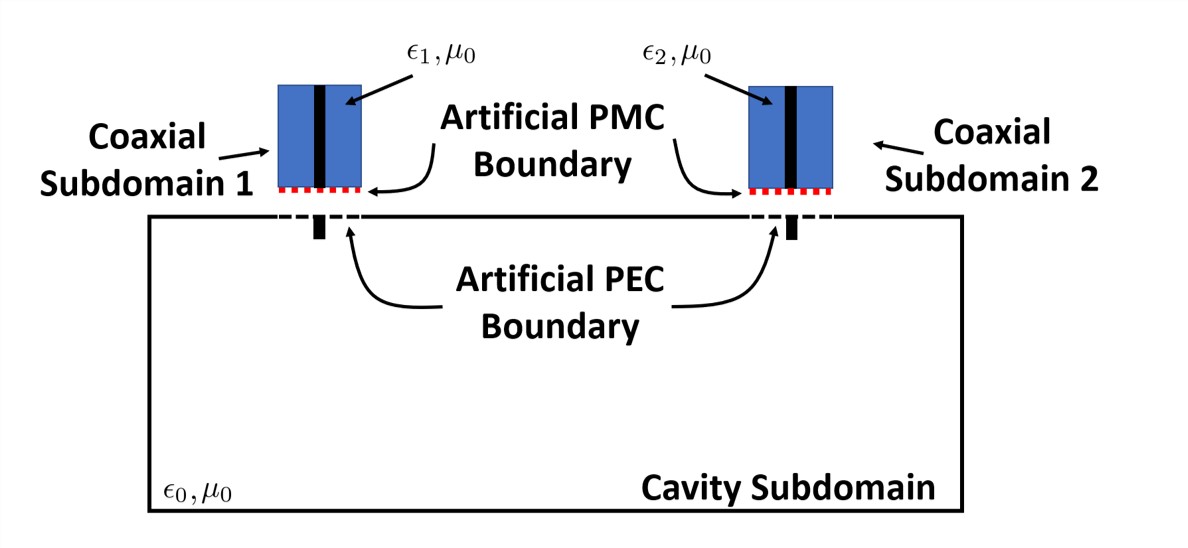}
		\caption{}
		\label{subfig:geometry_empty_system_modematching}
    \end{subfigure}
    \caption{Illustration of the subdomains for the mode-matching field quantization. In (a), the reference planes and subdomains are identified, while in (b) artificial boundaries have been introduced to separate the subdomain eigenvalue problems.}
    \label{fig:geometry_empty_system}
\end{figure}

Here, we consider an empty cavity that is fed by multiple waveguiding ports that are assumed to be semi-infinite in length while maintaining a constant cross-sectional shape. Systems with such semi-infinite ports can be most easily analyzed using a \textit{mode-matching (or projector-based) field quantization} method \cite{roth2021macroscopic,viviescas2003field} that enables domain decomposition concepts to be rigorously used in the quantization of the system. The basic process of this domain decomposition is illustrated in Fig. \ref{fig:geometry_empty_system} for a waveguide cavity fed by two coaxial ports.

In the mode-matching quantization approach, the system is divided into smaller subdomains so that each can be treated as a separate eigenvalue problem. To separate the subdomains, artificial boundary conditions are introduced at the interface of each set of subdomains. To maintain the hermiticity of the subdomain eigenvalue problems, complementary perfect electric conductor (PEC) and perfect magnetic conductor (PMC) conditions are assigned on either side of the subdomain interface, as in Fig. \ref{subfig:geometry_empty_system_modematching}. A complete set of orthonormal eigenmodes can now be found in each subdomain that are then quantized. To recover the complete system description, the dynamics of the quantum operators in the subdomains are tied together through interactions facilitated at the subdomain interfaces. Physically, these interactions can be viewed as being due to the interplay between equivalent electric or magnetic currents expanded in terms of the eigenmodes from one subdomain and the magnetic or electric vector potentials expanded in terms of the eigenmodes from the other subdomain \cite{roth2021macroscopic}.

Now, we consider the mathematical description of this procedure for the system shown in Fig. \ref{subfig:geometry_empty_system_modematching} that closes the waveguide cavity region with PEC boundary conditions and all ports with PMC boundary conditions. In each subdomain, the electric field operator $\hat{\mathbf{E}}$ must satisfy the wave equation
\begin{align}
    \nabla \times \nabla \times \hat{\mathbf{E}}(\mathbf{r},t) + \mu_0 \epsilon \partial^{2}_{t} \hat{\mathbf{E}}(\mathbf{r},t) = 0,
    \label{eq:waveequation}
\end{align}
where we assume for simplicity that a spatially-varying permittivity can occur but that there are no magnetic materials present. This wave equation can be solved using a separation of variables approach to decompose the field operator into a spatial part that is a vector field eigenmode and a temporal part that is a quantum operator. In the cavity subdomain, the electric and magnetic field operators can then be expressed using an eigenmode expansion as 
\begin{align}
    \hat{\mathbf{E}}_c(\mathbf{r},t)  = \sum_{k}^{} \sqrt{\frac{\omega_{k}}{\epsilon_{0}}}\hat{q}_k(t) {\mathbf{E}}_{k}(\mathbf{r}),
    \label{eq:eigenmode expansion - Efield - cavity}
\end{align}
\begin{align}
    \hat{\mathbf{H}}_c(\mathbf{r},t)  = \sum_{k}^{} \sqrt{\frac{\omega_{k}}{\mu_{0}}}\hat{p}_k(t)\mathbf{H}_{k}(\mathbf{r}),
    \label{eq:eigenmode expansion - Hfield - cavity}
\end{align}
where $\mathbf{E}_k(\mathbf{r})$ and $\mathbf{H}_k(\mathbf{r})$ are the spatial eigenmode of the electric field and magnetic field associated with the eigenvalue $\omega_k$, and $\hat{q}_k$ and $\hat{p}_{k}$ are canonically conjugate Hermitian quantum operators with commutation relation $[\hat{q}_{k_1},\hat{p}_{k_2}] = i\hbar\delta_{k_1k_2}$. Note $k$ here is an integer index unrelated to the EM wavenumber. The spatial eigenmodes are orthonormal in the sense of
\begin{align}
    \iiint \epsilon_r(\mathbf{r}) \mathbf{E}_{k_1}(\mathbf{r}) \cdot \mathbf{E}_{k_2}(\mathbf{r}) d\mathbf{r} = \delta_{k_1 k_2},
\end{align}
with a similar relation holding for the $\mathbf{H}_k$'s as well. Typically, the operators $\hat{q}_k$ and $\hat{p}_k$ are combined to form bosonic annihliation and creation operators for the $k$th field mode as
\begin{align}
    \hat{a}_k(t) = \frac{1}{\sqrt{2\hbar}} (\hat{q}_k(t) + i\hat{p}_k(t)),
    \label{eq:canonical_conjugate_field_var_a}
\end{align}
\begin{align}
    \hat{a}^\dagger_k(t) = \frac{1}{\sqrt{2\hbar}} (\hat{q}_k(t) - i\hat{p}_k(t)),
    \label{eq:canonical_conjugate_field_var_a_dagger}
\end{align}
respectively. It can be easily shown that these operators satisfy the commutation relation $[\hat{a}_{k_1},\hat{a}^\dagger_{k_2}] = \delta_{k_1 k_2}$. Using these operators, we can rewrite (\ref{eq:eigenmode expansion - Efield - cavity}) and (\ref{eq:eigenmode expansion - Hfield - cavity}) as
\begin{align}
    \hat{\mathbf{E}}_{c}(\mathbf{r},t)  = \sum_{k}^{} \sqrt{\frac{\hbar\omega_{k}}{2\epsilon_{0}}} \big(\hat{a}_k(t) + \hat{a}^{\dag}_k(t) \big) \mathbf{E}_{k}(\mathbf{r}),
    \label{eq:eigenmode expansion - Efield - cavity1}
\end{align}
\begin{align}
    \hat{\mathbf{H}}_{c}(\mathbf{r},t)  = -i\sum_{k}^{} \sqrt{\frac{\hbar\omega_{k}}{2\mu_{0}}}\big(\hat{a}_k(t) - \hat{a}^{\dag}_k(t) \big)\mathbf{H}_{k}(\mathbf{r}).
    \label{eq:eigenmode expansion - Hfield - cavity1}
\end{align}

For the fields in the $p$th coaxial port subdomain, there is a continuum of modes due to its semi-infinite length so the expansion becomes
\begin{multline}
    \hat{\mathbf{E}}_{p}(\mathbf{r},t)  =  \sum_\lambda \int_{0}^{\infty} \sqrt{\frac{\hbar\omega_{\lambda p}}{2\epsilon_{0}}}  \big(\hat{a}_{\lambda p}(\omega_{\lambda p},t)\! +\! \hat{a}^{\dag}_{\lambda p}(\omega_{\lambda p},t)\big) \\ \times\mathbf{E}_{\lambda p}(\omega_{\lambda p},\mathbf{r}) \, d\omega_{\lambda p},
    \label{eq:eigenmode expansion - Efield - port}
\end{multline}
\begin{multline}
    \hat{\mathbf{H}}_{p}(\mathbf{r},t)  =  -i\sum_\lambda \int_{0}^{\infty} \sqrt{\frac{\hbar\omega_{\lambda p}}{2\mu_{0}}}  \big(\hat{a}_{\lambda p}(\omega_{\lambda p},t)\! -\! \hat{a}^{\dag}_{\lambda p}(\omega_{\lambda p},t)\big) \\ \times\mathbf{H}_{\lambda p}(\omega_{\lambda p},\mathbf{r}) \, d\omega_{\lambda p},
    \label{eq:eigenmode expansion - Hfield - port}
\end{multline}
where $p$ indexes the coaxial subdomains and $\lambda$ differentiates between transverse mode profiles with corresponding eigenvalues $\omega_{\lambda p}$. Further, $\hat{a}_{\lambda p}$ and $\hat{a}^\dagger_{\lambda p}$ have similar relationships to those defined in (\ref{eq:canonical_conjugate_field_var_a}) and (\ref{eq:canonical_conjugate_field_var_a_dagger}). In the continuum case, these operators have commutation relation
\begin{multline}
    [\hat{a}_{\lambda_1 p_1}(\omega_{\lambda_1 p_1},t) , \hat{a}^\dagger_{\lambda_2 p_2}(\omega'_{\lambda_2 p_2},t)] = \delta_{\lambda_1 \lambda_2} \delta_{p_1 p_2} \\ \times \delta(\omega_{\lambda_1 p_1} - \omega'_{\lambda_2 p_2}).
\end{multline}

We can now consider the complete Hamiltonian for the system in Fig. \ref{fig:geometry_empty_system}. This characterizes the total energy in the system, which for this case is given by
\begin{align}
    \hat{H}_{tot,CP} = \hat{H}_C + \hat{H}_P +\hat{H}_{CP},
    \label{eq:total_Hamiltonian_empty}
\end{align}
where 
\begin{align}
    \hat{H}_C = \frac{1}{2} \iiint \big( \epsilon\hat{\mathbf{E}}^{2}_c + \mu_0\hat{\mathbf{H}}^{2}_c \big)dV
    \label{eq:Hamiltonian_Cavity_field}
\end{align}
is the EM energy integrated over the cavity subdomain,
\begin{align}
\hat{H}_P = \sum_p \frac{1}{2} \iiint \big( \epsilon_p\hat{\mathbf{E}}^{2}_p + \mu_0\hat{\mathbf{H}}^{2}_p \big)dV
\label{eq:Hamiltonian_Port_field}
\end{align}
is the EM energy integrated over the port subdomains, and 
\begin{align}
    \hat{H}_{CP} = - \sum_p \iint \hat{\mathbf{F}}_c \cdot \big(  \hat{\mathbf{E}}_{ p} \times \tilde{n}_p  \big)  dS
    \label{eq:Hamiltonian_cavity_port_interaction}
\end{align}
corresponds to the interaction Hamiltonian describing the coupling between the cavity and port fields. In (\ref{eq:Hamiltonian_cavity_port_interaction}), the surface integration occurs over the interface between subdomains, $\tilde{n}_p$ is the unit normal pointing into the cavity, and
\begin{align}
    \hat{\mathbf{F}}_c(\mathbf{r},t) = -\sum_k \sqrt{\frac{\hbar}{2\omega_k \mu_0}} \big( \hat{a}_k(t) + \hat{a}^\dagger_k(t) \big) \mathbf{H}_k(\mathbf{r}) 
\end{align}
is the electric vector potential in the cavity region.

Using the eigenmode orthonormality, the spatial integrals in (\ref{eq:total_Hamiltonian_empty}) can be evaluated. Performing this, $\hat{H}$ simplifies to
\begin{multline}
    \hat{H}_{tot,CP} = \sum_k \hbar\omega_k \hat{a}^\dagger_k \hat{a}_k \\ + \sum_{\lambda,p} \int_0^\infty \hbar \omega_{\lambda p} \hat{a}^\dagger_{\lambda p} (\omega_{\lambda p}) \hat{a}_{\lambda p}(\omega_{\lambda p}) d\omega_{\lambda p} + \\
     \sum_{k, p , \lambda } \int_{0}^{\infty}  \hbar g_{k,\lambda p}  \big(\hat{a}_{k} \!+\! \hat{a}_{k}^{\dag} \big) \big(\hat{a}_{\lambda p}(\omega_{\lambda p}) \!+ \! \hat{a}_{\lambda p}^{\dag}(\omega_{\lambda p}) \big)  d \omega_{\lambda p} ,
     \label{eq:Hamiltonian_empty_cavity_simplified}
\end{multline}
where 
\begin{align}
    g_{k,\lambda p} = \iint \frac{c_0}{2}\sqrt{\frac{\omega_{\lambda p}}{\omega_k}} \big[\mathbf{H}_{k} \cdot (\mathbf{E}_{\lambda p} \times \tilde{n}_p) \big] dS.
    \label{eq:coupling-strength}
\end{align}

This Hamiltonian formalism has been validated in the classical regime in \cite{moon2023full,moon2023validation} by computing scattering parameters and comparing to traditional finite element simulations. This also validates this quantum Hamiltonian since the field-based aspects of it are identical to the classical case. In Section \ref{sec:empty-cavity-analytical}, equations of motion will be derived from (\ref{eq:Hamiltonian_empty_cavity_simplified}) and subsequently solved using quantum input-output theory to develop an analytical solution for the port-fed empty cavity case.

\subsection{Closed Cavity and Qubit}
\label{subsec:background-cavity+qubit}
\begin{figure}[t!]
    \centering
    \includegraphics[width=0.8\linewidth]{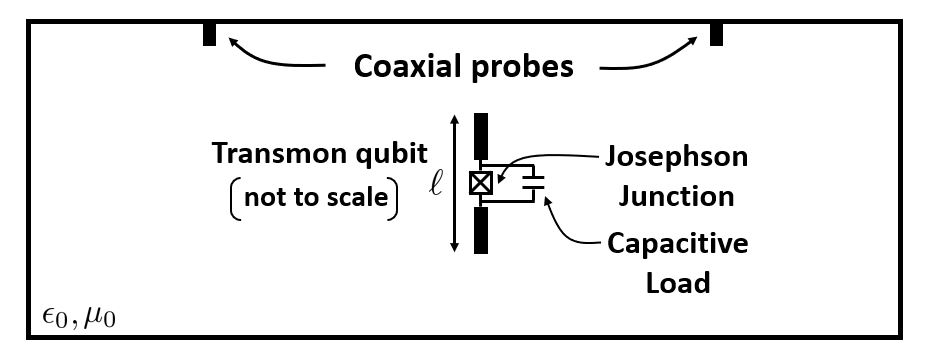}
    \caption{Schematic illustration of a transmon embedded in a rectangular waveguide cavity structure that is analytically solvable for all field-based aspects of the quantum full-wave Hamiltonian analysis.}
    \label{fig:geometry_qubit_system}
\end{figure}

Next, we consider the case with a transmon qubit embedded into a closed cavity as shown in Fig. \ref{fig:geometry_qubit_system}. Here, the transmon qubit is formed by a Josephson junction and additional capacitive load connected across the terminals of a small linear dipole antenna. To keep our later analytical solution more amenable to extensions, we retain the probes in our closed region that would connect to coaxial ports, although these do not serve a strictly necessary purpose here and can be omitted if desired. The total Hamiltonian for Fig. \ref{fig:geometry_qubit_system} is then
\begin{align}
\hat{H}_{tot,CT} = \hat{H}_C + \hat{H}_T + \hat{H}_{CT},
\label{eq:total_Hamiltonian_transmon}
\end{align}
where $\hat{H}_C$ is the cavity energy given in (\ref{eq:Hamiltonian_Cavity_field}), $\hat{H}_T$ is the transmon energy, and $\hat{H}_{CT}$ is the interaction energy due to the coupling of the transmon and cavity fields.

More specifically, the transmon Hamiltonian is
\begin{align}
\hat{H}_T = 4E_C\hat{n}^2 - E_J\cos\hat{\varphi},
\label{eq:transmon_Hamiltonian}
\end{align}
which from a circuit theory perspective corresponds to a linear capacitor in parallel with a nonlinear inductor. Here, $\hat{n}$ and $\hat{\varphi}$ are the qubit charge and phase operators that serve as (dimensionless) canonically conjugate operators for the qubit \cite{koch2007charge,roth2022transmon}. Further, $E_C = e^2/(2C_{\Sigma})$ is the charging energy of the total qubit capacitance $C_{\Sigma}$, where $e$ is the electron charge. We also have that the Josephson energy is $E_J = (\hbar/2e)^2/L_J$, where $L_J$ is the Josephon inductance. For a transmon, the energy parameters are designed such that $E_J/E_C \gg 1$ to minimize the qubit sensitivty to a common form of noise \cite{koch2007charge,roth2022transmon}. Meanwhile, the interaction Hamiltonian is 
\begin{align}
\hat{H}_{CT} = 2e\int\hat{\mathbf{E}}_c\cdot\mathbf{d}(\mathbf{r})\hat{n} d\mathbf{r},
\label{eq:Hamiltonian_int_CT}
\end{align}
where $\mathbf{d}$ parameterizes a line integration path so that its integral with $\hat{\mathbf{E}}_c$ computes the voltage seen by the Josephson junction in the transmon qubit \cite{roth2021macroscopic}. In the case of Fig. \ref{fig:geometry_qubit_system}, this would correspond to the voltage induced across the terminals of the dipole forming the transmon qubit. If there is more than one qubit in the cavity, one simply sums over independent qubits in the expressions of (\ref{eq:transmon_Hamiltonian}) and (\ref{eq:Hamiltonian_int_CT}) to generalize the mathematical description to this case.

Now, we can simplify (\ref{eq:total_Hamiltonian_transmon}) by substituting in the cavity field expressions from (\ref{eq:eigenmode expansion - Efield - cavity1}) and (\ref{eq:eigenmode expansion - Hfield - cavity1}) and re-expressing the transmon operators in terms of the energy eigenstates of (\ref{eq:transmon_Hamiltonian}) denoted by $|j\rangle$ in Dirac bra-ket notation. Performing this, we arrive at
\begin{multline}
\hat{H}_{tot,CT}  = \sum_{k} \hbar\omega_{k} {\hat{a}}^\dag_k {\hat{a}}_k  +
\sum_{j} \hbar \omega_{j} |{j}\rangle\langle{j}| \\+
\sum_{k,j} \big(\hbar{g_{k,j}\hat{a}}^\dag_k  |j\rangle \langle j+1|  + \hbar{g_{k,j}^{*} \hat{a}}_k |j+1\rangle \langle j| \big),
\label{eq:total_Hamiltonian2}
\end{multline}
where $\omega_j$ is the energy eigenvalue of (\ref{eq:transmon_Hamiltonian}) and we have also applied the rotating wave approximation to the interaction terms of $\hat{H}_{CT}$ \cite{koch2007charge}. Further, $g_{k,j}$ is the coupling rate between specific field and transmon modes given by
\begin{align}
g_{k,j} =2e \langle j | \hat{n} | j+1 \rangle  \sqrt{\frac{\omega_{k}}{2\epsilon_{0}\hbar} }\int{\mathbf{E}_k(\mathbf{r}})\cdot \mathbf{d}(\mathbf{r}) d\mathbf{r}.
\label{eq:coupling_strength_g}
\end{align}
It should be noted that we have also applied the approximation that the charge operator $\hat{n}$ only allows transitions between nearest-neighbor energy eigenstates for transmon qubits \cite{koch2007charge}. We will utilize this approximation to simplify our analytical solution; however, if the more general interactions are needed these can be computed analytically as well using the details of \cite{koch2007charge} or numerically using a simple method like in \cite{roth2023finite}.

\section{Port-Fed Empty Cavity Analytical Solution}
\label{sec:empty-cavity-analytical}
In this section, we develop an analytical solution for an empty cavity fed by ports. We discuss in Section \ref{subsec:quantum-io-theory} how to apply quantum input-output theory to the Hamiltonian of Section \ref{subsec:background-w-ports}. This results in a set of transfer functions that we can compute using analytical methods from EM theory. We then discuss in Section \ref{subsec:hom-results} how to use these transfer functions to compute HOM interference curves. We present results in Section \ref{subsec:empty-cavity-results} to validate the analytical solution.

\subsection{Quantum Input-Output Theory}
\label{subsec:quantum-io-theory}
Before beginning the analysis, it will be necessary to simplify the mode-matching Hamiltonian developed in Section \ref{subsec:background-w-ports} to make it analytically tractable \cite{walls2007quantum}. We will apply the rotating wave approximation and restrict our analysis to only consider a single cavity mode (with annihilation operator $\hat{a}_0$) whose resonance frequency we wish to compute transfer functions nearby (here we will consider the fundamental cavity mode) and the continuum of $\mathrm{TEM}$ modes in the coaxial ports (with annihilation operators $\hat{a}_p$). Applying these simplifications, the total Hamiltonian in (\ref{eq:Hamiltonian_empty_cavity_simplified}) reduces to
\begin{multline}
    \hat{H} \!=\!  \hbar\omega_0 \hat{a}^\dagger_0 \hat{a}_0 + \sum_{p} \int_0^\infty \hbar \omega_{p} \hat{a}^\dagger_{ p} (\omega_{ p}) \hat{a}_{ p}(\omega_{ p}) d\omega_{ p} \,+ \\
     \sum_{ p } \int_{0}^{\infty} \!\! \hbar g_{ p}  \big(\hat{a}_{0}^{\dag}\hat{a}_{ p}(\omega_{ p}) \!+\! \hat{a}_{0} \hat{a}_{ p}^{\dag}(\omega_{ p}) \big)  d \omega_{ p},
     \label{eq:Hamiltonian_empty_cavity_simplified_RWA}
\end{multline}
where $g_p$ is the restriction of (\ref{eq:coupling-strength}) to the modes specified above (the index $\lambda$ has been omitted as it no longer serves a purpose).

Now, to begin the analysis, we need the equations of motion for the cavity and port field operators. An equation of motion for an arbitrary quantum operator $\hat{X}$ can be found in the Heisenberg picture of quantum mechanics as \cite{miller2008quantum}
\begin{align}
    \partial_t \hat{X}  =  \frac{1}{i\hbar}[\hat{X},\hat{H}].
    \label{eq:EoM_definition}
\end{align}
Using the commutators for the annihilation and creation operators presented in Section \ref{subsec:background-w-ports}, we can determine that the equations of motion for $\hat{a}_0$ and an $\hat{a}_p$ are
\begin{align}
    \partial_t \hat{a}_{0}(t)  =  -i\omega_{0}\hat{a}_{0}(t) - i \sum_{p}\int_{-\infty}^{\infty}  g_{p} \hat{a}_{p}(t,\omega_p)  d\omega_{p} ,
	\label{eq:cavity-eom}
\end{align}
\begin{align}
	\partial_t \hat{a}_{p}(t,\omega_p) = -i\omega_{p}\hat{a}_{p}(t,\omega_p)  -i  g_{p} \hat{a}_0(t).
	\label{eq:coaxial-eom}
\end{align}
In deriving these, we have also assumed that the port fields are narrowband relative to their center frequency such that the frequency integration range can be extended from 0 to $-\infty$ (this will allow Fourier theory to be used later) \cite{walls2007quantum}.

The next step in quantum input-output theory is to integrate the equations of motion in the port subdomains in terms of initial and final conditions, which are taken to be well before and well after the interaction with the cavity has occurred so that they can be considered to be the ``input'' and ``output'' fields, respectively. Considering this, the integration of the port subdomain equations yields
\begin{multline}
	\hat{a}_{p}(t,\omega_p) = e^{-i\omega_p(t-t_0)}\hat{a}_{p}(t_0,\omega_p) \\-ig_{p}  \int_{t_0}^{t} e^{-i\omega_p(t-t^{\prime})}\hat{a}_{0}(t^{\prime})dt^{\prime}, \,\,\, \mathrm{for} \,\, t > t_0,
    \label{eq:coax_solution_input_condition}
 \end{multline}
 \begin{multline}
	\hat{a}_{p}(t,\omega_p) = e^{-i\omega_p(t-t_1)}\hat{a}_{p}(t_1,\omega_p) \\+ig_{p}  \int_{t}^{t_1} e^{-i\omega_p(t-t^{\prime})}\hat{a}_{0}(t^{\prime})dt^{\prime}, \,\,\, \mathrm{for} \,\, t < t_1,
    \label{eq:coax_solution_output_condition}
 \end{multline}
where $t_0$ ($t_1$) is the initial (final) condition time. These can then be substituted into (\ref{eq:cavity-eom}) to get
\begin{multline}
    \partial_t \hat{a}_{0}(t)  =  -i\omega_{0}\hat{a}_{0}(t) \\ - i \sum_{p} \int_{-\infty}^{\infty} g_{p} e^{-i\omega_p(t-t_0)}\hat{a}_{p}(t_0,\omega_p)  d\omega_{p} \\-  \sum_{p}\int_{-\infty}^{\infty}  g^{2}_{p} \int_{t_0}^{t} e^{-i\omega_p(t-t^{\prime})}\hat{a}_{0}(t^{\prime})dt^{\prime}  d\omega_{p},
	\label{eq:cavity-eom-intermediate-eq1}
\end{multline}
\begin{multline}
    \partial_t \hat{a}_{0}(t)  =  -i\omega_{0}\hat{a}_{0}(t) \\ - i \sum_{p} \int_{-\infty}^{\infty}  g_{p} e^{-i\omega_p(t-t_1)}\hat{a}_{p}(t_1,\omega_p)  d\omega_{p} \\+ \sum_{p}\int_{-\infty}^{\infty}   g^{2}_{p} \int_{t}^{t_1} e^{-i\omega_p(t-t^{\prime})}\hat{a}_{0}(t^{\prime})dt^{\prime}  d\omega_{p}.
	\label{eq:cavity-eom-intermediate-eq2}
\end{multline}

Next, we can apply standard Fourier transform identities to simplify the final terms in (\ref{eq:cavity-eom-intermediate-eq1}) and (\ref{eq:cavity-eom-intermediate-eq2}). To do this, we first make the Markov approximation by assuming that $g_p$ varies slowly enough over the frequency range of interest (nominally, the cavity resonance bandwidth) so that it can be factored out of the frequency integrals \cite{walls2007quantum}. Doing this and switching the order of integration in the final term of (\ref{eq:cavity-eom-intermediate-eq1}), we get
\begin{multline}
   \int_{-\infty}^{\infty}  g^{2}_{p} \int_{t_0}^{t} e^{-i\omega_p(t-t^{\prime})}\hat{a}_{0}(t^{\prime})dt^{\prime}  d\omega_{p} \\ \approx  g^{2}_{p}(\omega_0) \int_{t_0}^{t} \bigg[\int_{-\infty}^{\infty}   e^{-i\omega_p(t-t^{\prime})} d\omega_{p} \bigg] \hat{a}_{0}(t^{\prime}) dt^{\prime}.
	\label{eq:cavity-eom-intermediate-eq1-lastterm1}
\end{multline}
Noting the term inside the brackets equals $2\pi\delta(t-t')$, we get
\begin{multline}
    \int_{-\infty}^{\infty}  g^{2}_{p} \int_{t_0}^{t} e^{-i\omega_p(t-t^{\prime})}\hat{a}_{0}(t^{\prime})dt^{\prime}  d\omega_{p} \\ \approx 2\pi g^{2}_{p}(\omega_0) \int_{t_0}^{t}  \delta(t - t^{\prime}) \hat{a}_{0}(t^{\prime}) dt^{\prime}.  
	\label{eq:cavity-eom-intermediate-eq1-lastterm3}
\end{multline}
Since the integration range in (\ref{eq:cavity-eom-intermediate-eq1-lastterm3}) only covers half of the Dirac delta, the Fourier theory result of
\begin{multline}
    \int_{t_0}^{t}  \delta(t - t^{\prime}) \hat{a}_{0}(t^{\prime}) dt^{\prime} = \int_{t}^{t_1}  \delta(t - t^{\prime}) \hat{a}_{0}(t^{\prime}) dt^{\prime} \\= \frac{1}{2}\hat{a}_{0}(t^{\prime}) \,\,\, \mathrm{for} \,\, t_0 < t < t_1
	\label{eq:cavity-eom-intermediate-eq1-lastterm5}
\end{multline}
can be used (where the identity involving $t_1$ is used to simplify the synonymous term in (\ref{eq:cavity-eom-intermediate-eq2})) \cite{walls2007quantum}. Combining these simplifications, we finally have that (\ref{eq:cavity-eom-intermediate-eq1}) and (\ref{eq:cavity-eom-intermediate-eq2}) become
\begin{multline}
    \partial_t \hat{a}_{0}(t) = -i\omega_0\hat{a}_0(t) \\ -\sum_{p}\big(i\sqrt{2\pi} g_{p} \hat{a}_{\mathrm{in},p}(t)+\pi g^{2}_{p}\hat{a}_0(t)\big),
    \label{eq:cavity-eom-ver1}
\end{multline}
\begin{multline}
    \partial_t \hat{a}_{0}(t) = -i\omega_0\hat{a}_0(t) \\ -\sum_{p}\big(i\sqrt{2\pi} g_{p} \hat{a}_{\mathrm{out},p}(t)-\pi g^{2}_{p}\hat{a}_0(t)\big)
    \label{eq:cavity-eom-ver2}
\end{multline}
where we have simplified the notation by defining $\hat{a}_{\mathrm{in},p}(t)$ and $\hat{a}_{\mathrm{out},p}(t)$ as
\begin{align}
    \hat{a}_{\mathrm{in},p}(t) = \frac{1}{\sqrt{2 \pi}}\int_{-\infty}^{\infty}  e^{-i\omega_p(t-t_0)}\hat{a}_{p}(t_0,\omega_p)  d\omega_{p},
	\label{eq:ain}
\end{align}
\begin{align}
    \hat{a}_{\mathrm{out},p}(t)  =   \frac{1}{\sqrt{2 \pi}}\int_{-\infty}^{\infty}   e^{-i\omega_p(t-t_1)}a_{p}(t_1,\omega_p)  d\omega_{p}.
	\label{eq:aout}
\end{align}

At this point, it is convenient to exploit the linearity of the system to perform a time-harmonic analysis. Taking the Fourier transform of (\ref{eq:cavity-eom-ver1}) and (\ref{eq:cavity-eom-ver2}), we arrive at
\begin{multline}
	-i\omega{ \hat{a}_{0}(\omega)} = -i\omega_0\hat{a}_0(\omega)  \\ -\sum_{p}^{}(i\sqrt{2\pi}g_{p}\hat{a}_{\mathrm{in},p}(\omega)+\pi g^{2}_{p}\hat{a}_0(\omega)),
    \label{eq:cavity-eom-ver1-fourier}
\end{multline}
\begin{multline}
	-i\omega{ \hat{a}_{0}(\omega)} = -i\omega_0\hat{a}_0(\omega)  \\ -\sum_{p}^{}(i\sqrt{2\pi}g_{p}\hat{a}_{\mathrm{out},p}(\omega)-\pi g^{2}_{p}\hat{a}_0(\omega)).
    \label{eq:cavity-eom-ver1-fourier2}
\end{multline}
Subtracting (\ref{eq:cavity-eom-ver1-fourier2}) from (\ref{eq:cavity-eom-ver1-fourier}) results in what is referred to as a ``boundary condition'' in quantum input-output theory at each port \cite{walls2007quantum}, which in the frequency domain is
\begin{align}
	\hat{a}_{\mathrm{in},p}(\omega) - \hat{a}_{\mathrm{out},p}(\omega) = i\sqrt{2\pi}g_{p}\hat{a}_{0}(\omega).
 \label{eq:coax-cavity-boundary-condition}
\end{align}

To find the cavity transfer functions within this formalism, we first eliminate the cavity mode in (\ref{eq:cavity-eom-ver1-fourier}) by substituting into this equation using (\ref{eq:coax-cavity-boundary-condition}) evaluated for the first port. This yields
\begin{align}
    \hat{a}_{\mathrm{out},1}(\omega) = R_1(\omega) \hat{a}_{\mathrm{in},1}(\omega) + T_{12}(\omega) \hat{a}_{\mathrm{in},2}(\omega),
    \label{eq:aout1_final}
\end{align}
\begin{align}
    R_1(\omega) = \frac{\pi(g_2^2-g_1^2)-i(\omega-\omega_0)}{\pi(g_1^2+g_2^2) - i(\omega-\omega_0)},
\end{align}
\begin{align}
    T_{12}(\omega) = -\frac{2\pi g_1g_2}{\pi(g_1^2+g_2^2) - i(\omega-\omega_0)}.
\end{align}
A similar procedure for the second port yields
\begin{align}
    \hat{a}_{\mathrm{out},2}(\omega) = R_2(\omega) \hat{a}_{\mathrm{in},2}(\omega) + T_{21}(\omega) \hat{a}_{\mathrm{in},1}(\omega),
    \label{eq:aout2_final}
\end{align}
\begin{align}
    R_2(\omega) = \frac{\pi(g_1^2-g_2^2)-i(\omega-\omega_0)}{\pi(g_1^2+g_2^2) - i(\omega-\omega_0)},
\end{align}
and $T_{21}(\omega) = T_{12}(\omega)$. These correspond to Lorentzian transfer functions, which match what would be expected for a classical analysis \cite{moon2023validation}. However, it must be noted that in the quantum domain it is impossible to ``turn off'' the input at one of the ports and retain a valid quantum description. This is a result of vacuum fluctuations, which have important consequences for describing the correct behavior of such a system when considering input states that are not ``sufficiently classical'' (such as few-photon inputs) \cite{gerry2005introductory}. We will use these transfer functions in Section \ref{subsec:hom-results} to compute HOM interference curves, which require this correct quantum treatment \cite{na2020classical}.

We now turn to evaluating these transfer functions using analytical methods from EM theory. For simplicity, we will keep the length of the coaxial probes protruding into the cavity to be small so that their impact can be accounted for using microwave cavity perturbation theory for ``shape perturbations'' \cite{pozar2009microwave}. In this case, the resonant frequency of the dominant cavity mode including the effect of the perturbations is evaluated as
\begin{align}
\omega_{0} \approx \omega_{0}^{\prime} + \omega_{0}^{\prime}\,\frac{\displaystyle \iiint_{\Delta V}(\mu|\mathbf{H}_0|^2 - \epsilon|\mathbf{E}_0|^2)dV}{\displaystyle \iiint_{V}(\mu|\mathbf{H}_0|^2 + \epsilon|\mathbf{E}_0|^2)dV},
\label{eq:cavity_perturb_theory}
\end{align}
where $\omega_{0}$ is the perturbed resonant frequency, $\omega_{0}^{\prime}$ is the unperturbed resonant frequency (i.e., of the empty cavity), $V$ is the volume of the unperturbed cavity, $\Delta V$ is the volume of the perturbation corresponding to the coaxial probes, $\mathbf{E}_0$ and $\mathbf{H}_0$ are the EM fields of the unperturbed cavity for the dominant field mode, and $\mu$ and $\epsilon$ are the constitutive parameters of the material inside the cavity (free space in the case considered here). Here, we integrate the EM energy over the coaxial probes by sampling the EM field at the tips of the coaxial probes and multiplying by the volume of the probe \cite{pozar2009microwave}, which is reasonable for the dominant cavity mode as it does not vary along the length of the coaxial probes. 

We can also use the basic concept of cavity perturbation theory to evaluate $g_1$ and $g_2$, however, more care is needed here on the model settings for this approximation to be reasonable. In particular, we assume in the evaluation of (\ref{eq:coupling-strength}) that we can use the unperturbed magnetic field $\mathbf{H}_0$. Near the probe, the magnetic field is obviously perturbed and the true field mode $\mathbf{H}_{k}$ will deviate from the unperturbed field distribution. However, we have found this change to the field distribution of $\mathbf{H}_{k}$ to be very localized to the region immediately surrounding the probe. As a result, if the integration region $dS$ is taken over a reasonably large radius the error from the perturbation theory assumption can be kept at a tolerable level. Sample settings to achieve this will be shown in Section \ref{subsec:empty-cavity-results}.

\subsection{Hong-Ou-Mandel Interference}
\label{subsec:hom-results}
The HOM effect is a quantum optical phenomena commonly used to test the quality of single photon sources \cite{lounis2005single,scheel2009single}. In a HOM experiment, two different input EM fields are directed at a 50/50 ``beam splitter'' that produces a $90^\circ$ phase shift between reflected and transmitted fields. If these EM fields correspond to indistinguishable single photons (i.e., they have identical properties except that they are incident on the beam splitter from different ports/directions), then a \textit{photon bunching} effect occurs wherein the two photons will always exit one of the two output ports of the beam splitter \textit{together}. To experimentally observe this effect, photon counting detectors are placed in the two output ports of the beam splitter and coincidences (i.e., when both photon detectors trigger) are recorded. To generate a HOM interference curve, these experiments are repeated with different delays between the incident photons. If the incident photons are not indistinguishable or the beam splitter does not possess the necessary ideal properties, the quality of the HOM interference curve will degrade.

Mathematically, a HOM interference curve corresponds to calculating the second-order correlation function given by \cite{gerry2005introductory}
\begin{multline}
    g^{(2)}(\tau) = \frac{A}{BC} = \\ \frac{ \langle \psi | \hat{E}^{(-)}_1(t_0) \hat{E}^{(-)}_2(t_0\!+\!\tau)  \hat{E}^{(+)}_2(t_0\!+\!\tau) \hat{E}^{(+)}_1(t_0)  | \psi \rangle}{ \langle \psi | \hat{E}^{(-)}_1(t_0) \hat{E}^{(+)}_1(t_0) | \psi \rangle  \langle \psi | \hat{E}^{(-)}_2(t_0\!+\!\tau)  \hat{E}^{(+)}_2(t_0\!+\!\tau)  | \psi \rangle  },
\end{multline}
where $\tau$ is the delay between incident photons, $\hat{E}^{(\pm)}_j(t)$ is the positive (negative) frequency component of the electric field operator in output port $j$ evaluated at time $t$ and $|\psi\rangle$ is the initial joint quantum state of the system. If the two input photons are indistinguishable, then $g^{(2)}(0) = 0$. In general, a value of $g^{(2)}(\tau) < 0.5$ is taken to be a signature of a quantum EM field source, as typical classical states of EM fields cannot achieve these values \cite{kaltenbaek2006experimental,rarity2005non,sadana2019near}. 

Now, we specialize our discussion to how to use the results of Section \ref{subsec:quantum-io-theory} to compute a HOM interference curve. In this case, our cavity acts as the ``beam splitter'' where the two coaxial ports serve both as input and output ports. The electric field operators in the output ports are then proportional to the output creation and annihilation operators as
\begin{align}
    \hat{E}^{(+)}_j(t) \propto \hat{a}_{\mathrm{out},j}(t), \,\,\,\, \hat{E}^{(-)}_j(t) \propto \hat{a}^\dagger_{\mathrm{out},j}(t),
\end{align}
where the remaining proportionality constants will cancel in the evaluation of $g^{(2)}$ and so are omitted for simplicity. It is now necessary to express $\hat{a}_{\mathrm{out},j}(t)$ in terms of the input bosonic operators using (\ref{eq:aout1_final}) and (\ref{eq:aout2_final}) since the quantum state $|\psi\rangle$ will be specified in terms of the initial conditions of the input photon states. Further, we must take the inverse Fourier transform of these expressions to write our operators in the time domain to evaluate the $g^{(2)}$ function. Again dropping constant terms that will cancel, we can approximate the inverse Fourier transform through a discrete summation as
\begin{multline}
    \hat{a}_{\mathrm{out},1}(t) = \sum_{m} \bigg[ R_1(\omega_m) \hat{a}_{\mathrm{in},1}(\omega_m) \\ + T_{12}(\omega_m) \hat{a}_{\mathrm{in},2}(\omega_m) \bigg] e^{-i\omega_m t},
\end{multline}
\begin{multline}
    \hat{a}^\dagger_{\mathrm{out},1}(t) = \sum_{m} \bigg[ R_1^*(\omega_m) \hat{a}^\dagger_{\mathrm{in},1}(\omega_m) \\ + T^*_{12}(\omega_m) \hat{a}^\dagger_{\mathrm{in},2}(\omega_m) \bigg] e^{i\omega_m t},
\end{multline}
with a synonymous set of results also for the second port.

Finally, we must specify the initial state of the system $|\psi\rangle$ to be able to compute the $g^{(2)}$ function. Here, we will only consider unentangled single photon states in both input ports, but other input states can also be analyzed (e.g., multi-photon or thermal states). To describe a single photon state, we operate on the vacuum state of a particular port with a linear expansion of the input creation operators from that port. Considering this, the joint input state is given by
\begin{multline}
    |\psi\rangle = \bigg( \sum_m W_{\mathrm{in},1}(\omega_m) \, \hat{a}^\dagger_{\mathrm{in},1}(\omega_m)  \bigg) |0\rangle_1 \\ \otimes \bigg( \sum_n W_{\mathrm{in},2}(\omega_n) \,\hat{a}^\dagger_{\mathrm{in},2}(\omega_n) \bigg) |0\rangle_2,
    \label{eq:io-input-state}
\end{multline}
where $|0\rangle_j$ is the vacuum state in input port $j$ and $W_{\mathrm{in},j}(\omega_m)$ is a corresponding spectral weight that defines the temporal shape of the input photons. These spectral weighting coefficients can be easily computed from the Fourier transform of the desired temporal shape of the single photons. Here we will take these to be modulated Gaussian functions, so that the spectral weights are
\begin{align}
    W_{\mathrm{in},1}(\omega) = \frac{1}{\sqrt{N}} e^{-(\sigma_1 (\omega-\omega_{\mathrm{in},1}))^2/2} \, e^{i \omega t_0},
\end{align}
\begin{align}
    W_{\mathrm{in},2}(\omega) = \frac{1}{\sqrt{N}}e^{-(\sigma_2 (\omega-\omega_{\mathrm{in},2}))^2/2} \, e^{i \omega (t_0+\tau)},
\end{align}
where $\sigma_j$ is the temporal standard deviation of the modulated Gaussian pulse in port $j$ and $\omega_{\mathrm{in},j}$ is the corresponding center frequency, and $N$ is a normalization constant so that the overall quantum state in (\ref{eq:io-input-state}) is normalized.

\begin{figure*}[!t]
\normalsize
\newcounter{MYtempeqncnt}
\newcounter{MYtempeqncnt2}
\setcounter{MYtempeqncnt}{\value{equation}}
\setcounter{equation}{54}
\begin{multline}
    A \!=\! \sum_{m,n} \bigg( W^*_{\mathrm{in},1}(\omega_m) R^*_1(\omega_m) e^{i\omega_m t_0} \, W^*_{\mathrm{in},2}(\omega_n) R^*_2(\omega_n) e^{i\omega_n(t_0+\tau)} + W^*_{\mathrm{in},1}(\omega_m) T^*_{21}(\omega_m) e^{i\omega_m (t_0+\tau)} \, W^*_{\mathrm{in},2}(\omega_n) T^*_{12}(\omega_n) e^{i\omega_n t_0}  \bigg) \\ \times \sum_{\ell,p} \bigg( W_{\mathrm{in},1}(\omega_\ell) R_1(\omega_\ell) e^{-i\omega_\ell t_0} \, W_{\mathrm{in},2}(\omega_p) R_2(\omega_p) e^{-i\omega_p(t_0+\tau)} + W_{\mathrm{in},1}(\omega_p) T_{21}(\omega_p) e^{-i\omega_p (t_0+\tau)} \, W_{\mathrm{in},2}(\omega_\ell) T_{12}(\omega_\ell) e^{-i\omega_\ell t_0}  \bigg)
    \label{eq:g2_a}
\end{multline}
\begin{multline}
    B = \bigg( \sum_m W_{\mathrm{in},2}(\omega_m) T_{12}(\omega_m) e^{-i\omega_m t_0}  \bigg)  \bigg( \sum_n W^*_{\mathrm{in},2}(\omega_n) T^*_{12}(\omega_n) e^{i\omega_n t_0} \bigg) \bigg( \sum_\ell | W_{\mathrm{in},1}(\omega_\ell)  |^2 \bigg) \\ + \bigg( \sum_m W_{\mathrm{in},1}(\omega_m) R_{1}(\omega_m) e^{-i\omega_m t_0}  \bigg)   \bigg( \sum_n W^*_{\mathrm{in},1}(\omega_n) R^*_{1}(\omega_n) e^{i\omega_n t_0} \bigg) \bigg( \sum_\ell | W_{\mathrm{in},2}(\omega_\ell)  |^2 \bigg)
    \label{eq:g2_b}
\end{multline}
\begin{multline}
    C = \bigg( \sum_m W_{\mathrm{in},1}(\omega_m) T_{21}(\omega_m) e^{-i\omega_m (t_0+\tau)}  \bigg)  \bigg( \sum_n W^*_{\mathrm{in},1}(\omega_n) T^*_{21}(\omega_n) e^{i\omega_n (t_0+\tau)} \bigg) \bigg( \sum_\ell | W_{\mathrm{in},2}(\omega_\ell)  |^2 \bigg) \\ + \bigg( \sum_m W_{\mathrm{in},2}(\omega_m) R_{2}(\omega_m) e^{-i\omega_m (t_0+\tau)}  \bigg)   \bigg( \sum_n W^*_{\mathrm{in},2}(\omega_n) R^*_{2}(\omega_n) e^{i\omega_n (t_0+\tau)} \bigg) \bigg( \sum_\ell | W_{\mathrm{in},1}(\omega_\ell)  |^2 \bigg)
    \label{eq:g2_c}
\end{multline}
\hrulefill
\setcounter{MYtempeqncnt2}{\value{equation}}
\setcounter{equation}{\value{MYtempeqncnt}}
\end{figure*}

With all expressions appropriately defined, we can now evaluate the $g^{(2)}(\tau)$ function. The algebra of this is rather tedious, but after utilizing properties of the creation and annihilation operators we can find that the individual expressions simplify to (\ref{eq:g2_a}) to (\ref{eq:g2_c}), shown at the top of the next page. From these expressions, we can see that the only geometry-dependent parameters are the values of the transfer functions that were analytically computed using quantum input-output theory and cavity perturbation theory in Section \ref{subsec:quantum-io-theory}. Hence, we can now analytically calculate the HOM interference curves for this port-fed empty cavity system. 
\setcounter{equation}{\value{MYtempeqncnt2}}

\subsection{Results}
\label{subsec:empty-cavity-results}
To compute results, we consider a geometry as in Fig. \ref{fig:geometry_empty_system} that is composed of a rectangular waveguide cavity with two coaxial lines coupled to it. The dimension of the cavity is $22.86\times 10.16\times 40 \, \mathrm{mm}^3$. The coaxial probes are located along the center of the transverse dimension of the cavity, are both offset $10 \, \mathrm{mm}$ from the ends of the cavity along the longitudinal dimension, and have an inner conductor radius of $0.05 \, \mathrm{mm}$. Due to the symmetric positions of the coaxial probes and that the magnetic field of the dominant cavity mode is an odd function, we can conclude that $g_1 = -g_2$ for this system.

\begin{figure}[t]
    \centering
    \includegraphics[width=0.8\linewidth]{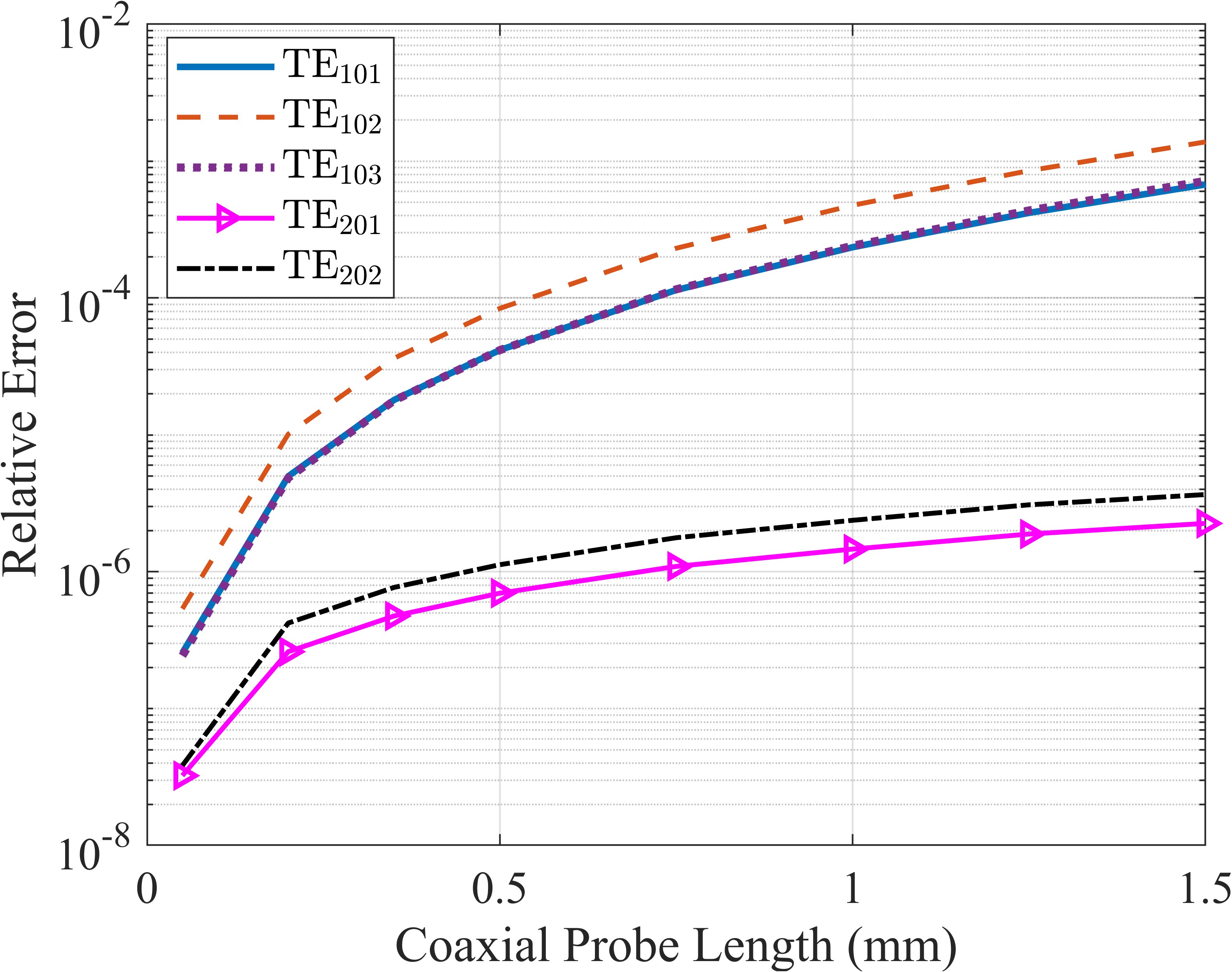}
    \caption{Relative error in the perturbation theory calculation of the cavity resonant frequency for the lowest five modes.}
    \label{fig:res-freq-rel-error}
\end{figure}

To begin validating our analytical formulation, we first consider the error in the perturbation theory when the length the inner conductor protrudes into the cavity is varied from $0.05 \, \mathrm{mm}$ to $1.5 \, \mathrm{mm}$. As a reference, we perform an eigenmode analysis of the cavity region including the perturbing coaxial inner conductors using the finite element method (FEM). To not bias the error calculation of the perturbation theory approximation due to numerical errors from FEM, we use as our unperturbed resonant frequency the FEM-computed eigenvalue for the cavity with no coaxial probes present in (\ref{eq:cavity_perturb_theory}). We then compute the relative error between the perturbation theory and FEM resonant frequencies for the lowest five cavity modes as a function of coaxial probe length (although not needed here, higher-order modes will be included in the calculations in Section \ref{sec:qubit-analytical}). These results are shown in Fig. \ref{fig:res-freq-rel-error}, where it is seen that the relative error can be quite low for small perturbations. 

\begin{figure}[t]
    \centering
    \includegraphics[width=0.8\linewidth]{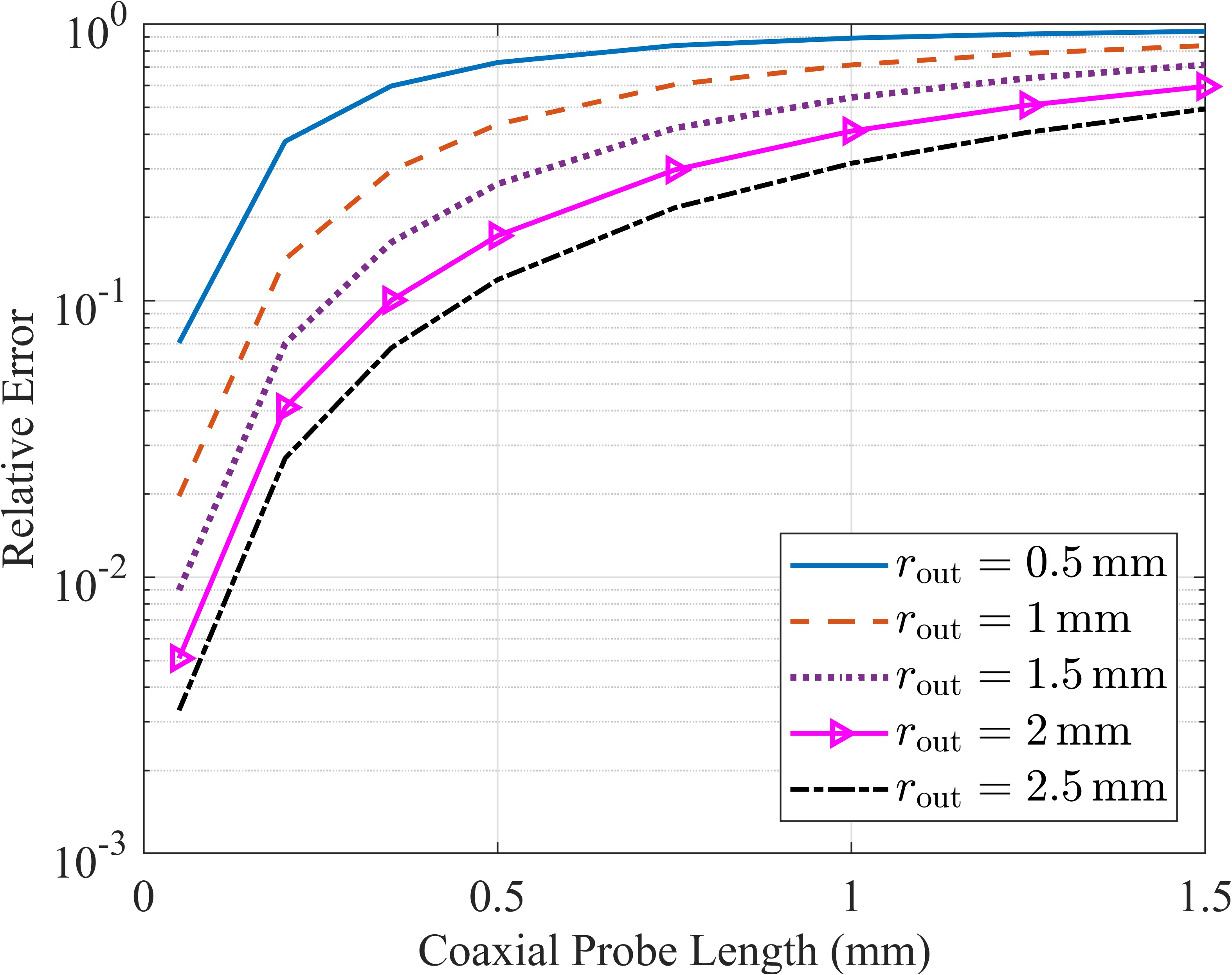}
    \caption{Relative error in the perturbation theory calculation of $g_j$ for the dominant cavity mode for different coaxial outer radii.}
    \label{fig:g-rel-error}
\end{figure}

Next, we perform a similar analysis, but for the computation of the $g_j$ value. This depends on the overlap integral between cavity and coaxial subdomains in the form of $\mathbf{H}_k\cdot \big( \mathbf{E}_{p} \times \tilde{n}_p  \big) $, where $\mathbf{H}_k$ is the cavity magnetic field mode and $\mathbf{E}_p$ is the TEM mode of the coaxial line. To evaluate this overlap integral, we apply a trapezoidal integration rule in the radial and azimuthal directions over an annular region centered around the inner conductor of the coaxial line. As mentioned earlier, the accuracy of the perturbation theory approximation can be improved if the surface the overlap integral is evaluated over is made larger (i.e., increasing the outer radius of the coaxial line). To show the impact of this, we compute the relative error in the calculation of $g_j$ when using perturbation theory or the eigenmodes from FEM in the cavity region for various coaxial probe lengths. As seen in Fig. \ref{fig:g-rel-error}, by careful selection of design parameters this error can also be kept small. 

\begin{figure}[t]
    \centering
    \begin{subfigure}[t]{0.85\linewidth}
        \includegraphics[width=\textwidth]{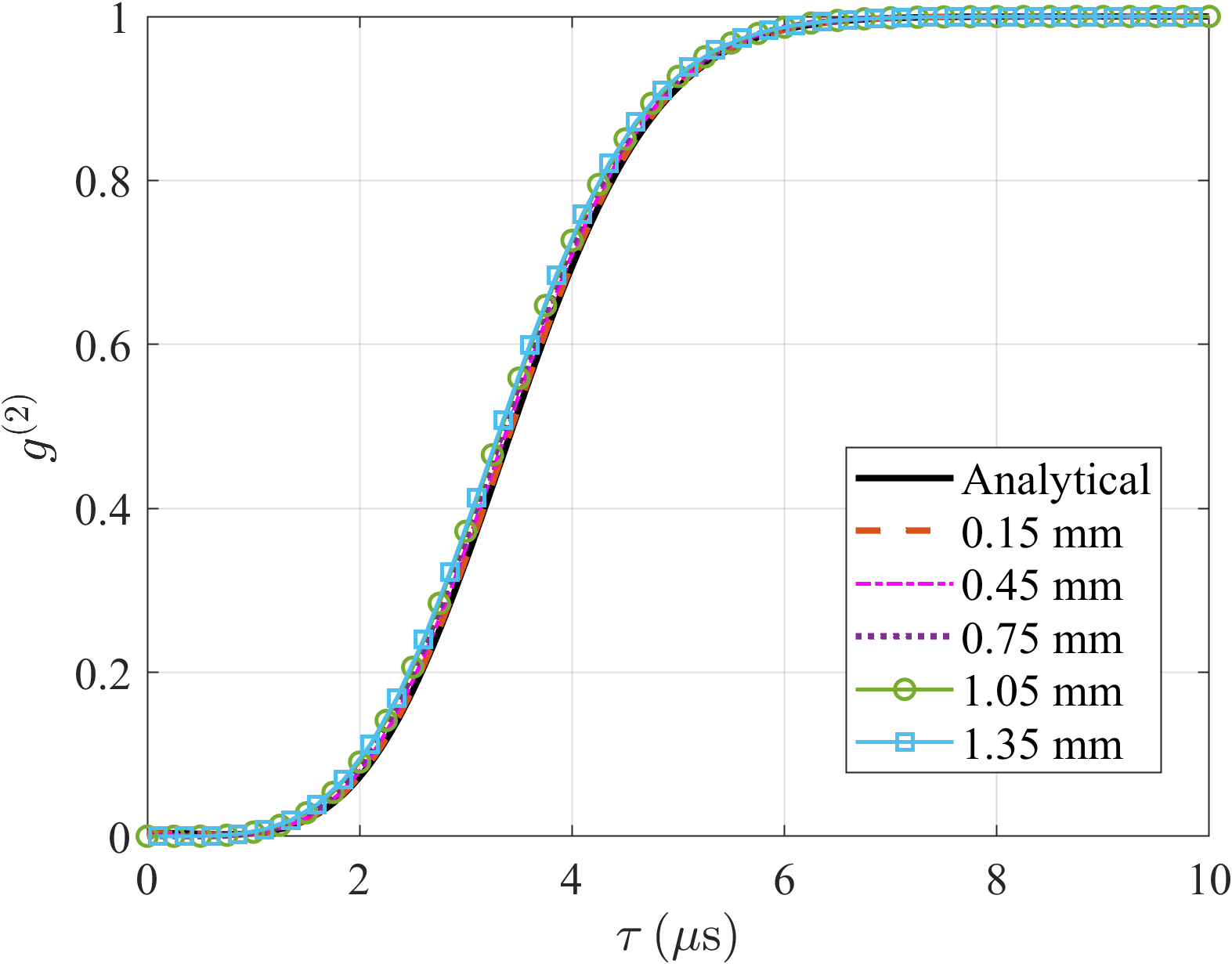}
        \caption{}
        \label{subfig:hom_s2p5e-6_optFreq}
    \end{subfigure}
    \begin{subfigure}[t]{0.85\linewidth}
		\includegraphics[width=\textwidth]{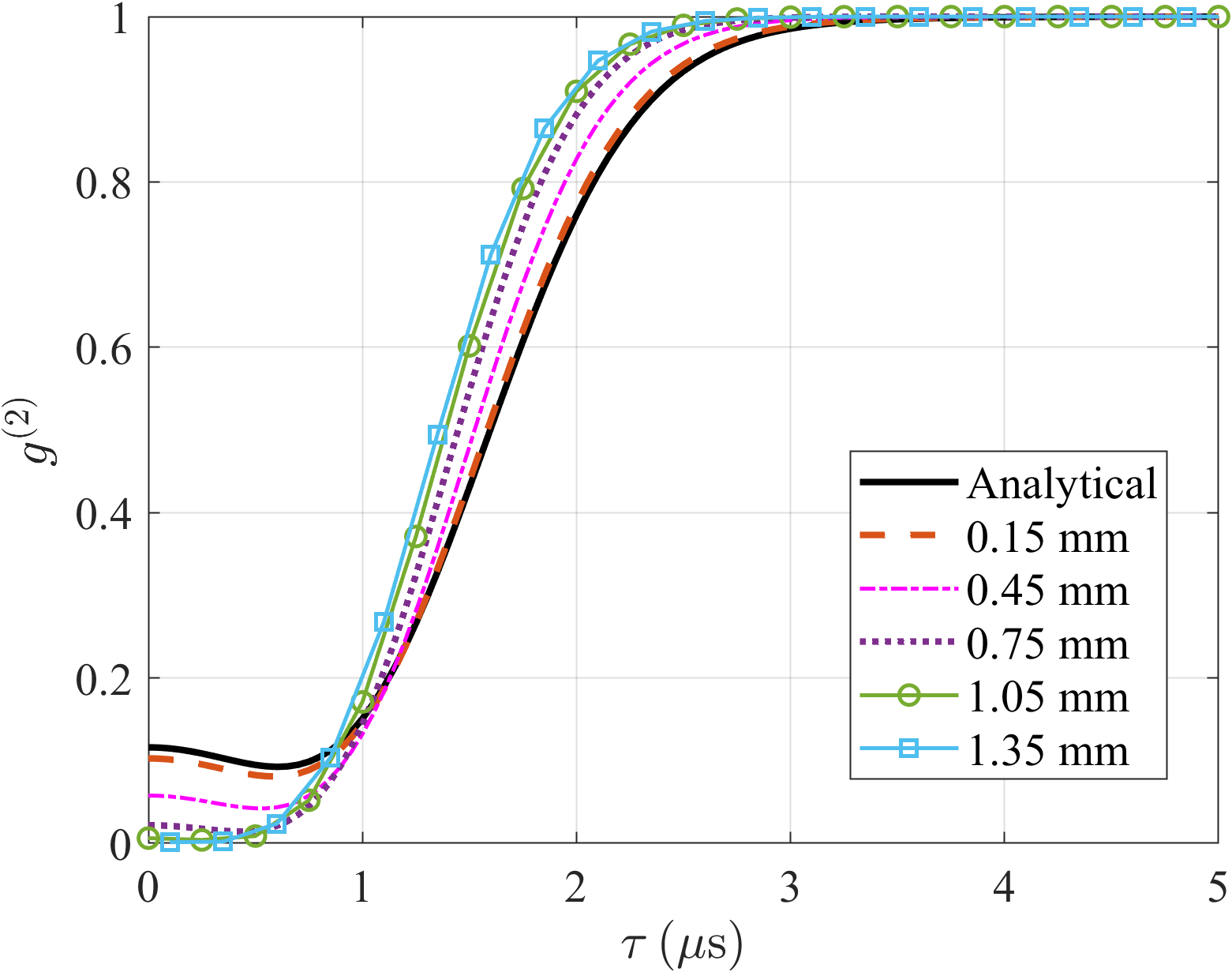}
		\caption{}
		\label{subfig:hom_s1e-6_optFreq}
    \end{subfigure}
    \caption{HOM interference curves with individually-optimized single photon center frequencies for analytical and numerical eigenmode calculations with different coaxial probe lengths and a coaxial outer radius of $2.5 \, \mathrm{mm}$. Temporal standard deviation of the Gaussian pulses are (a) $\sigma_1 = \sigma_2 = 2.5 \, \mu\mathrm{s}$ and  (b) $\sigma_1 = \sigma_2 = 1 \, \mu\mathrm{s}$.}
    \label{fig:hom_optFreq}
\end{figure}

We now show how these errors in the perturbation theory approach impact the calculation of a HOM interference curve for various scenarios. To begin, we isolate the impact of the error in the computation of $g_j$ by allowing the center frequencies of the single photons to be independently optimized for the Lorentzian transfer functions at various probe lengths for the analytical and numerical eigenmode computations when the coaxial outer radius is set to $2.5 \, \mathrm{mm}$. The results of this are shown in Fig. \ref{fig:hom_optFreq} for $\sigma_1 = \sigma_2 = 2.5 \, \mu\mathrm{s}$ and $1 \, \mu\mathrm{s}$. For the narrower photon bandwidth shown in Fig. \ref{subfig:hom_s2p5e-6_optFreq}, it is seen that the error in $g_j$ has a small impact on the overall HOM interference curve and that high-quality interference is always achieved. For the wider photon bandwidth shown in Fig. \ref{subfig:hom_s1e-6_optFreq}, the quality of the HOM curve is degraded for shorter probe lengths due to the higher $Q$ of the resonator. As the probe is made longer, the $Q$ lowers and a higher-quality HOM interference is then achieved again.

\begin{figure}[t]
    \centering
    \begin{subfigure}[t]{0.85\linewidth}
        \includegraphics[width=\textwidth]{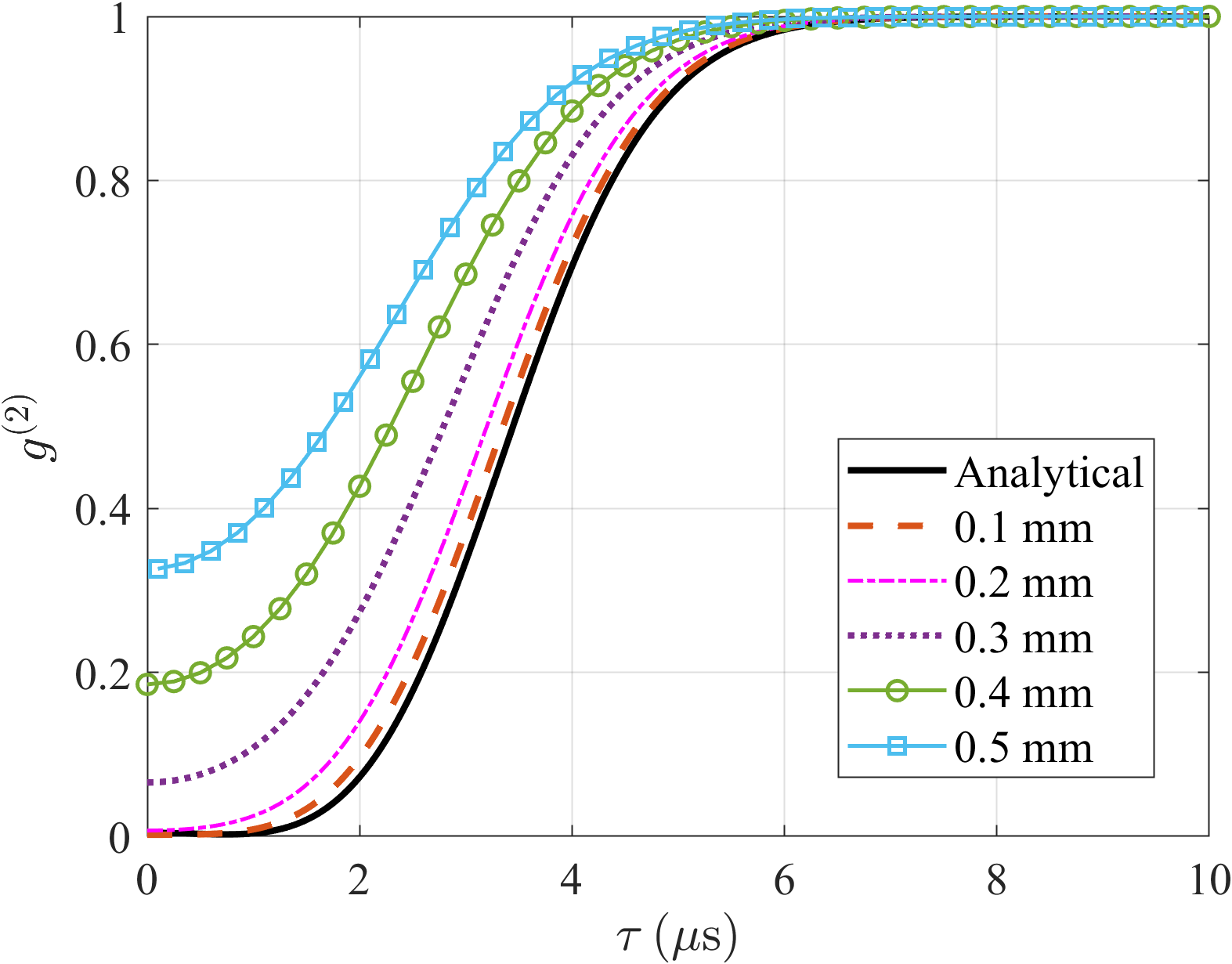}
        \caption{}
        \label{subfig:hom_s2p5e-6_fixedFreq}
    \end{subfigure}
    \begin{subfigure}[t]{0.85\linewidth}
		\includegraphics[width=\textwidth]{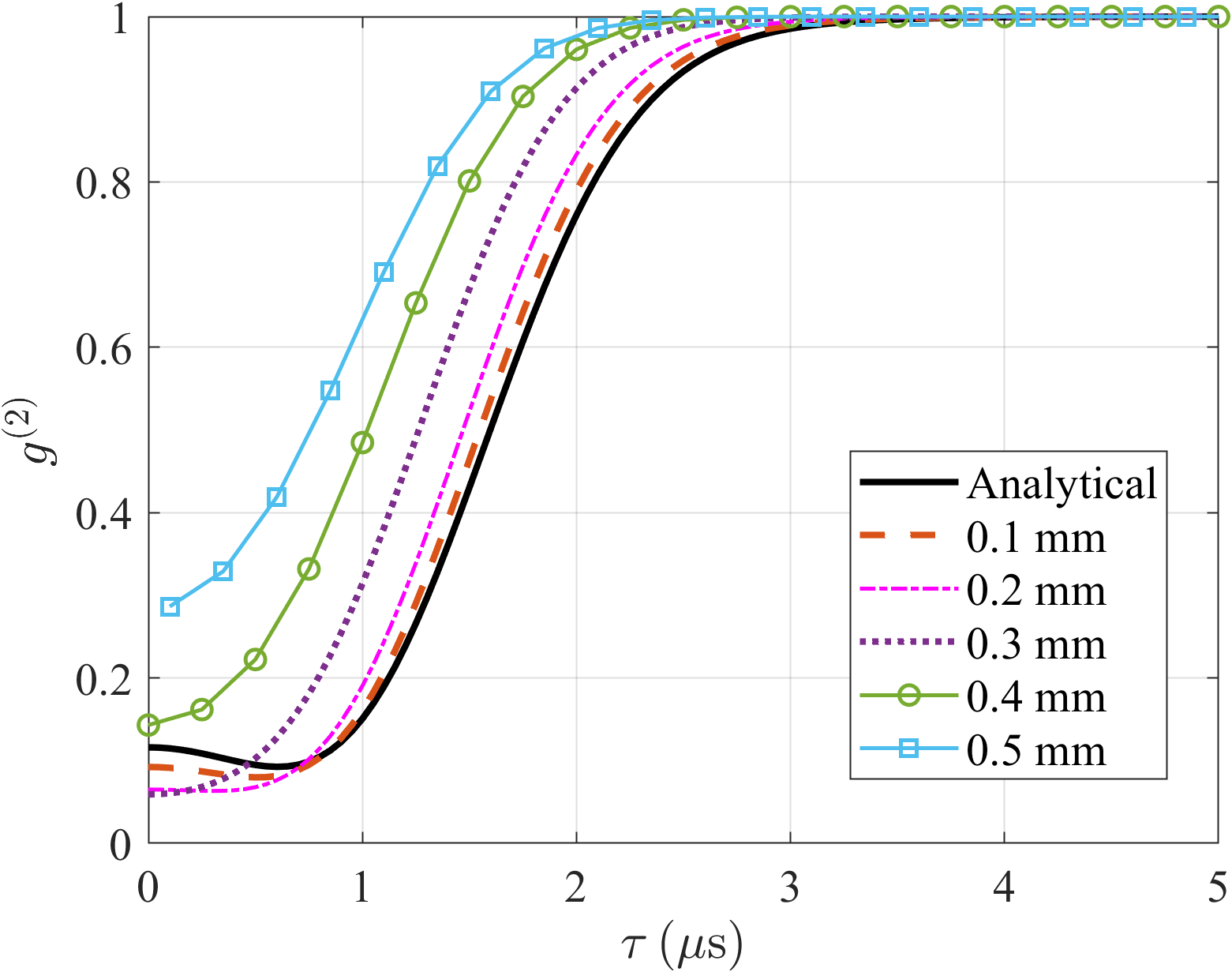}
		\caption{}
		\label{subfig:hom_s1e-6_fixedFreq}
    \end{subfigure}
    \caption{HOM interference curves with fixed single photon center frequencies for analytical and numerical eigenmode calculations with different coaxial probe lengths and a coaxial outer radius of $2.5 \, \mathrm{mm}$. Temporal standard deviation of the Gaussian pulses are (a) $\sigma_1 = \sigma_2 = 2.5 \, \mu\mathrm{s}$ and  (b) $\sigma_1 = \sigma_2 = 1 \, \mu\mathrm{s}$.}
    \label{fig:hom_fixedFreq}
\end{figure}

Next, we repeat these calculations but do not allow the center frequencies of the single photons to be independently optimized for the numerical eigenmode computations. Instead, we use fixed parameters based off of the transfer functions computed using perturbation theory. The results are shown in Fig. \ref{fig:hom_fixedFreq}, where it is seen that the error in the computation of the cavity resonant frequency quickly degrades the accuracy of the analytical solution as the probe length is increased. However, if the probe length is kept small, good agreement is still achievable. Overall, a more sophisticated classical EM analytical solution for the effect of the coaxial probe on the cavity field modes could extend the range of parameters our analytical quantum full-wave solution can be applied over.

\section{Closed Cavity and Qubit Analytical Solution}
\label{sec:qubit-analytical}
In this section, we develop an analytical solution for one or more transmon qubits inside a closed rectangular waveguide cavity. We discuss in Section \ref{subsec:qubit-analytical-parameters} how to use analytical methods from EM theory to evaluate the parameters comprising the Hamiltonian discussed in Section \ref{subsec:background-cavity+qubit}. Next, we discuss in Section \ref{subsec:dispersive-regime-calculations} how to utilize a matrix representation of the system Hamiltonian to compute experimentally-relevant system parameters that are important for the control and measurement of qubit states. We then present results in Section \ref{subsec:qubit-analytical-results} to validate our analytical solution.

\subsection{Analytical Evaluation of Hamiltonian Parameters}
\label{subsec:qubit-analytical-parameters}
As mentioned previously, the basic system geometry we will consider here corresponds to that shown in Fig. \ref{fig:geometry_qubit_system} which is inspired by 3D transmons that have been studied experimentally \cite{paik2011observation}. We have made the cavity a rectangular waveguide and include one or two transmons inside made from small linear dipole antennas so that typical EM and antenna theory techniques can be leveraged in our analytical solution \cite{balanis2016antenna}. In our system, we keep the length of the dipoles electrically small compared to the spatial variation of the relevant cavity field modes. This makes the cavity field mode profiles appear effectively like plane wavesfrom the perspective of the dipoles so that typical antenna theory formulas for free space operation can be applied in the cavity, as will be substantiated in Section \ref{subsec:qubit-analytical-results}. When considering two transmons, we keep the dipoles of different transmons located far enough away from each other to minimize mutual coupling that would invalidate approximations made in our analytical solution.

We now consider evaluating all the parameters in the total Hamiltonian of (\ref{eq:total_Hamiltonian2}). To begin, we will compute the cavity resonant frequency $\omega_k$ using cavity perturbation theory. Based on the field quantization process used for our analytical solution, we need to determine $\omega_k$ in the absence of the transmons, which leaves only the perturbation due to the coaxial probes to be accounted for. Given this, $\omega_k$ can be computed using the simple generalization of (\ref{eq:cavity_perturb_theory}) to the $k$th field mode.

Next, the eigenfrequencies of the transmons $\omega_j$ in (\ref{eq:total_Hamiltonian2}) will be considered. In typical scenarios, $E_J$ is a given parameter that depends on the microscopic structure of the Josephson junction and is not influenced by the surrounding geometry. Hence, the only other parameter needed to characterize a particular transmon is $E_C$, which is a function of the total capacitance in parallel to the Josephson junction. Here, this is 
\begin{align}
C_{\Sigma} = C_{ant} +  C_{L}
\label{eq:total-capacitance}
\end{align}
where $C_{ant}$ is the geometric capacitance of the dipole and $C_{L}$ is the total load capacitance due the linear Josephson junction capacitance and a lumped element placed across the dipole terminals. This additional lumped capacitance is needed to boost the total qubit capacitance so that the qubit operates in the ``transmon regime'' where $E_J/E_C \gg 1$. Since $C_L$ is a given parameter in a design, the only capacitance that needs to be determined from the geometry of the system is $C_{ant}$. For a small dipole, this capacitance can be computed using 
\begin{align}
C_{{ant}} = \frac{\tan(k\ell/2)}{120\omega_0(\ln(\ell/(2r))-1)},
\label{eq:balanis dipole antenna}
\end{align}
where $\ell$ is the length of the dipole, $r$ is the radius of the cylinder of the dipole, $k$ is the wavenumber of the homogeneous medium filling the cavity, and $\omega_0$ is the operating frequency of the fundamental mode of the cavity \cite{balanis2016antenna}. With $E_C$ and $E_J$ determined, they can be used to calculate the eigenfrequencies of the transmon either analytically \cite{koch2007charge} or numerically. In our work, we have used a simple finite element method to compute the eigenfrequencies of the transmon as described in \cite{roth2023finite}. Also, note that $C_{ant}$ computed from (\ref{eq:balanis dipole antenna}) is the main error source in our analytical solution. To improve the accuracy, a more sophisticated analytical formula for $C_{ant}$ could be developed or the capacitance could be estimated numerically from a classical simulation of a dipole in free space. 

Lastly, the coupling rate $g_{k,j}$ in (\ref{eq:coupling_strength_g}) needs to be evaluated. Beginning with the qubit-related properties, the transition matrix element of the charge operator $\langle j|\hat{n}|j+1 \rangle$ is needed. This can be computed numerically as in \cite{roth2023finite}, or can be approximated analytically as \cite{koch2007charge}
\begin{align}
\langle j|\hat{n}|j+1 \rangle \approx -i\bigg(\frac{E_J}{8E_C}\bigg)^{1/4}\sqrt{\frac{j+1}{2}}.
\label{eq:charge_operator}
\end{align}

\begin{figure}[t!]
    \centering
    \includegraphics[width=0.9\linewidth]{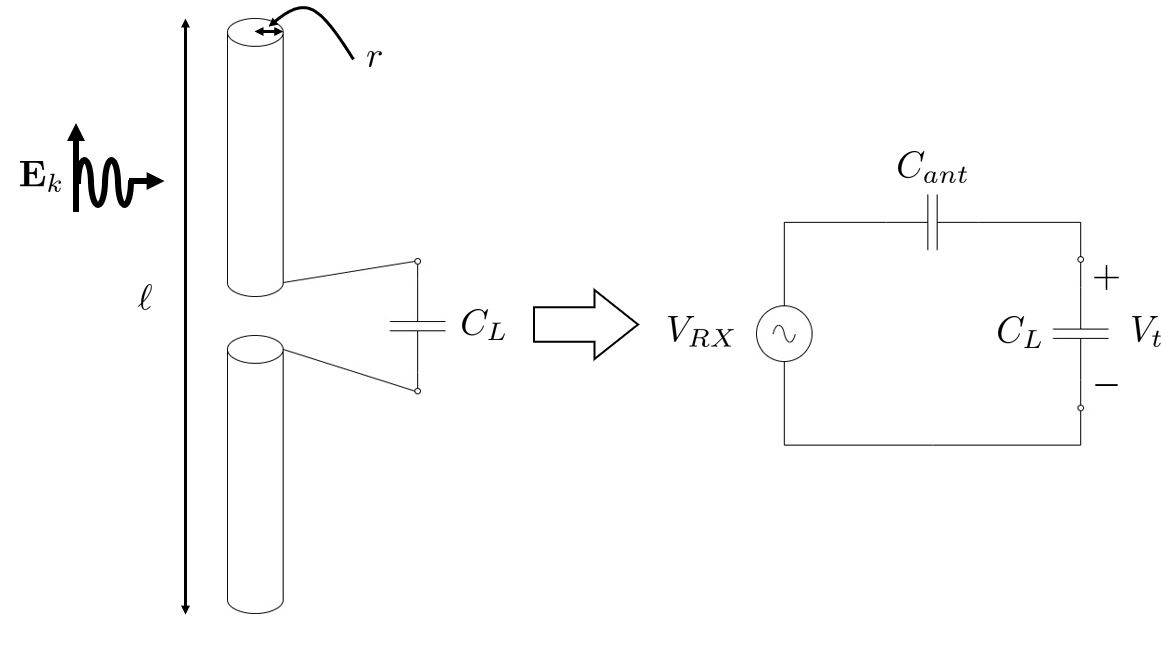}
    \caption{Schematic illustration of the Th\'{e}vinin equivalent circuit model of a receiving wire antenna for our case of a transmon formed with a load capacitance connected across the dipole terminals.}
    \label{fig:equivalent_circuit}
\end{figure}

The spatial integral in (\ref{eq:coupling_strength_g}) also must be evaluated, and corresponds to computing the voltage induced over the capacitance connected across the dipole terminals in the absence of the nonlinear inductance of the Josephson junction. This can be found using the theory of \cite{alu2010power, tretyakov2003analytical}, which rigorously establishes the Th\'{e}vinin equivalent circuit model of a receiving wire antenna with an arbitrary load, as shown in Fig. \ref{fig:equivalent_circuit}. In this approach, the amplitude of the equivalent voltage source is given by
\begin{align}
V_{RX} = \int_{-\ell/2}^{\ell/2} f_{TX}(\mathbf{r}) \mathbf{E}_{k}(\mathbf{r})\cdot \mathbf{d}(\mathbf{r}) d\mathbf{r},
\label{eq:induced_voltage}
\end{align}
where $\ell$ is the length of the dipole and $f_{TX}(\mathbf{r})$ is the normalized current distribution of the wire antenna in transmitting mode. For an electrically small dipole, the transmitting mode current distribution can be approximated as a triangular function \cite{balanis2016antenna}. If we assume that $\mathbf{E}_k$ does not vary over the length of the small dipole antenna, we can evaluate (\ref{eq:induced_voltage}) to get
\begin{align}
V_{{RX}} = \frac{1}{2} \, \hat{\ell}\cdot  \mathbf{E}_{k}(\mathbf{r}_0)\ell,
\label{eq:vector_effective_length}    
\end{align}
where $\hat{\ell}$ is a unit vector pointing along the length of the dipole and $\mathbf{r}_0$ is the position at the center of the dipole. Then, using the equivalent circuit in Fig. \ref{fig:equivalent_circuit}, the needed voltage induced across the load capacitance is simply
\begin{align}
V_{{t}} = \frac{C_{{ant}}}{C_{{ant}} + C_{L}} V_{{RX}}.
\label{eq:V_terminal_transmon}    
\end{align}

\subsection{Evaluation of Dispersive Regime System Parameters}
\label{subsec:dispersive-regime-calculations}
With all the parameters in the Hamiltonian operator now determined, a matrix representation of it can be found in terms of a suitable basis \cite{miller2008quantum}. This ``Hamiltonian matrix'' can then be used in various calculations to compute different parameters of interest. Here, we will focus on computing system parameters needed for cQED devices operated in the \textit{dispersive regime} of cavity quantum electrodynamics, which is the most common operating regime in practice \cite{gu2017microwave,krantz2019quantum,blais2021circuit}. This operating regime is achieved when the cavity resonant frequencies $\omega_k$ are strongly detuned from the qubit transition frequencies $\omega_{j,j+1} = \omega_{j+1}-\omega_j$ relative to the coupling strength $g_{k,j}$; i.e., $g_{k,j}/|\omega_k - \omega_{j,j'}| \ll 1$. When this is the case, there is low hybridization between the qubit and cavity modes that allows sufficient qubit controllability without resulting in excessive decay through effects like spontaneous emission \cite{houck2008controlling,roth2022full} or unwanted dynamics due to effects like vacuum Rabi oscillations \cite{walls2007quantum}.

Considering this, the particular experimentally-relevant parameters that we have computed to test our method are the first qubit transition frequency $\omega_{01}$, the qubit anharmonicity $\alpha = \omega_{12}-\omega_{01}$, the cavity resonant frequencies $\omega_k$, the AC-Stark shift $\chi$, and the $ZZ$-interaction rate $\zeta$ \cite{krantz2019quantum,blais2021circuit}. The AC-Stark shift characterizes the interaction strength between a specific pair of qubit and cavity modes in the dispersive regime, and is important for designing qubit state measurement protocols. The $ZZ$-interaction rate is similar, but describes the dispersive coupling strength between a pair of qubit modes, and is important for designing multi-qubit gates (but also contributes to deleterious quantum crosstalk effects). Each of these parameters can be computed in terms of low-lying eigenvalues of the total system Hamiltonian given in (\ref{eq:total_Hamiltonian2}), with example formulas given later in this section for clarity. 

To find these eigenvalues, we first compute the Hamiltonian matrix for a particular system using a basis composed of tensor product states between the ``natural'' basis of each of the constituent parts of the system. For the qubit, this corresponds to its free energy eigenstates that are solutions to (\ref{eq:transmon_Hamiltonian}). Similarly, each cavity mode is expressed in terms of its own eigenstates that correspond to a fixed number of photons in the mode (typically referred to as \textit{Fock states}). The matrix representation of $\hat{H}$ is then found by evaluating
\begin{align}
    H_{mn} = \langle m | \hat{H} |n \rangle,
\end{align}
where $H_{mn}$ is the element in the $m$th row and $n$th column of the Hamiltonian matrix and $|m\rangle$ and $|n\rangle$ are two states of the tensor product basis being used. 

Now, many of the dispersive regime parameters of interest can be immediately deduced from the eigenvalues of the Hamiltonian matrix, but the calculations of $\chi$ and $\zeta$ require slightly more processing. To be concrete in our discussion, assume that we are considering a simulation with two qubit and two cavity modes. In the dispersive regime, the low amount of mode hybridization makes it possible to use the same labeling of our original basis states for the eigenvectors of the coupled system. We can then denote a particular eigenvalue of the Hamiltonian matrix as $E_{ijk\ell}$, where all subscripts will take on integer values denoting the number of photons in a particular mode and the ordering is such that $i$ ($j$) corresponds to the first (second) qubit mode and $k$ ($\ell$) corresponds to the first (second) cavity mode. One can then inspect the effective dispersive regime Hamiltonian of the complete system to determine procedures for computing $\chi$ or $\zeta$ between particular sets of modes. For instance, to compute the AC-Stark shift due to the first qubit mode on the first cavity mode, we would need to populate both of these modes and then subtract out the energies due to the individual excitation energies. Mathematically, this is
\begin{align}
    \chi = E_{1010} - E_{1000} - E_{0010} - E_{0000},
\end{align}
where we also subtract out $E_{0000}$ so that our effective ground energy is zero. Similarly, we can compute the ZZ-interaction rate between the two qubit modes as
\begin{align}
    \zeta = E_{1100} - E_{1000} - E_{0100} - E_{0000}.
\end{align}
These formulas can be adjusted easily to compute the $\chi$ or $\zeta$ between other modes as needed.

In any calculation, it is imperative that a sufficiently large basis is used to achieve numerically converged results for the system parameters of interest. However, the size of the Hamiltonian matrix grows exponentially as the number of modes or number of quantum states per mode are increased. For instance, if the same number of quantum states is used irrespective of the type of mode (e.g., qubit or cavity), then the Hamiltonian matrix will have dimension of $M^N$ where $N$ is the number of modes and $M$ is the number of quantum states per mode. Due to this exponential growth, being able to achieve accurate numerical results with a lower value of $M$ is essential. As we will see in Section \ref{subsec:qubit-analytical-results}, our approach will be able to use a much smaller value of $M$ than the EPR method, leading to an exponentially smaller Hamiltonian matrix to achieve the same level of accuracy.

\subsection{Results}
\label{subsec:qubit-analytical-results}
To compute results, we consider a system geometry like that illustrated in Fig. \ref{fig:geometry_qubit_system} that is composed of a rectangular waveguide cavity with two coaxial probes and one or two transmon qubits. The dimensions of the cavity and coaxial probes are the same as those in Section \ref{subsec:hom-results} with the coaxial probe length fixed to a value of $0.75 \, \mathrm{mm}$. Each transmon consists of a linear dipole antenna with length $1 \, \mathrm{mm}$, radius $0.04 \, \mathrm{mm}$, terminal gap size of $0.102 \, \mathrm{mm}$, and is oriented along the electric field direction of the dominant cavity modes. A total load capacitance of $C_L = 50.34 \, \mathrm{fF}$ is also always used for each qubit. Further, the qubits are kept located at the central plane of the cavity along its smallest dimension.

\begin{figure}[t]
    \centering
    \begin{subfigure}[t]{0.49\linewidth}
        \includegraphics[width=\textwidth]{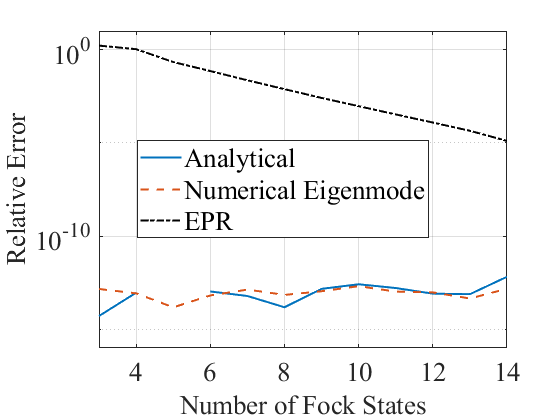}
        \caption{}
        \label{subfig:Qubit_anharmonicity_converegence_plot}
    \end{subfigure}
    \begin{subfigure}[t]{0.49\linewidth}
		\includegraphics[width=\textwidth]{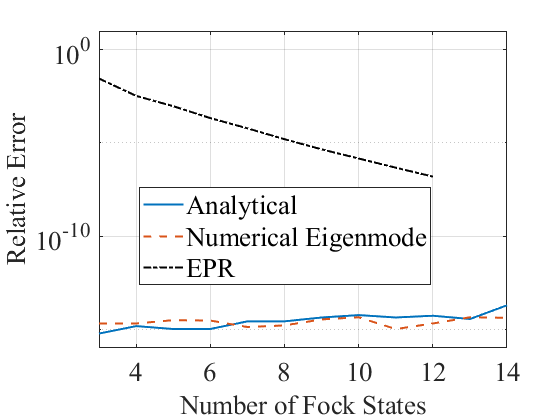}
		\caption{}
		\label{subfig:Qubit_Eigenfrequency1_converegence_plot}
    \end{subfigure}
    \caption{Numerical convergence of the different computation methods for (a) the qubit anharmonicity $\alpha$ and (b) the first qubit transition frequency $\omega_{01}$. The reference solution for each computation method is its own results using $15$ Fock states for each mode in the basis.}
    \label{fig:convergence_plots}
\end{figure}

As an initial test, we consider the numerical convergence of typical system parameters as a function of the number of basis states used for the qubit and cavity modes when there is only one qubit present, located at the center of the cavity, and only the first two cavity modes are included. For these tests, the Josephson junction of the transmon has $L_J = 9.4 \, \mathrm{nH}$. Convergence plots are shown in Fig. \ref{fig:convergence_plots} for $\alpha$ and $\omega_{01}$ computed using three different methods. In particular, for our analytical solution, our method but using numerical EM eigenmodes, and the EPR method. The relative error is computed for each method by comparing to the results of its own calculation using $15$ Fock states for each mode in the basis. From the results, it is clear that the field-based approach described here converges immediately with the lowest number of possible Fock states while the EPR method requires a substantial number of Fock states to converge.

We attribute our formulation's quick convergence to explicitly incorporating the nonlinearity of the qubit in the analysis of the qubit subsystem in formulating our Hamiltonian matrix. In contrast, the EPR method treats the nonlinear effects as a perturbation to the linear effects, which then requires a large number of Fock states to be considered to achieve convergence even in the properties of the lowest energy levels of the system. Since the size of the Hamiltonian matrix grows exponentially with the number of Fock states considered, achieving fast numerical convergence is a key property. 

\begin{figure}
    \centering
    \begin{subfigure}[t]{0.49\linewidth}
        \includegraphics[width=\textwidth]{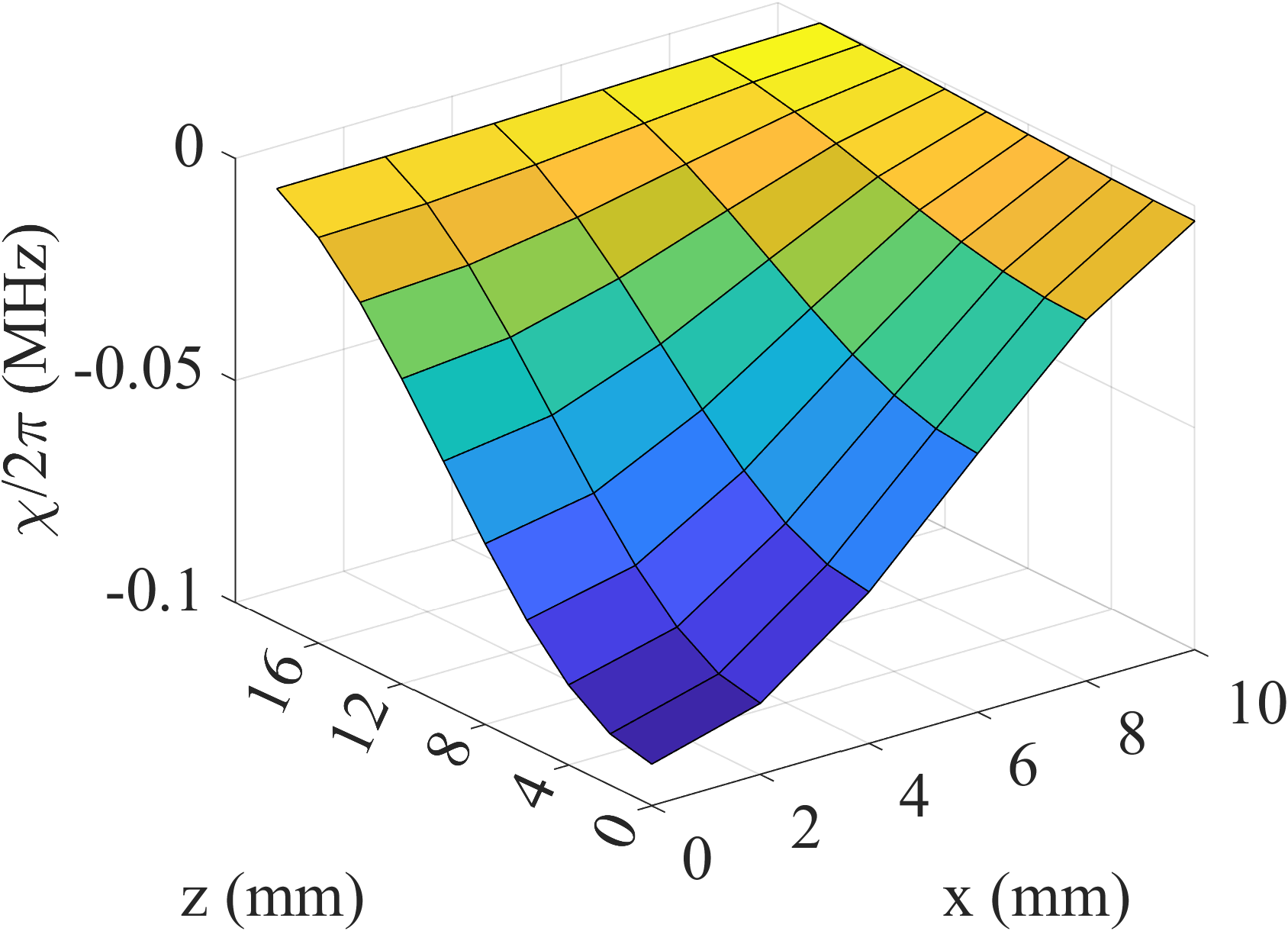}
        \caption{}
        \label{subfig:AC_stark_single_analytical}
    \end{subfigure}
    \begin{subfigure}[t]{0.49\linewidth}
		\includegraphics[width=\textwidth]{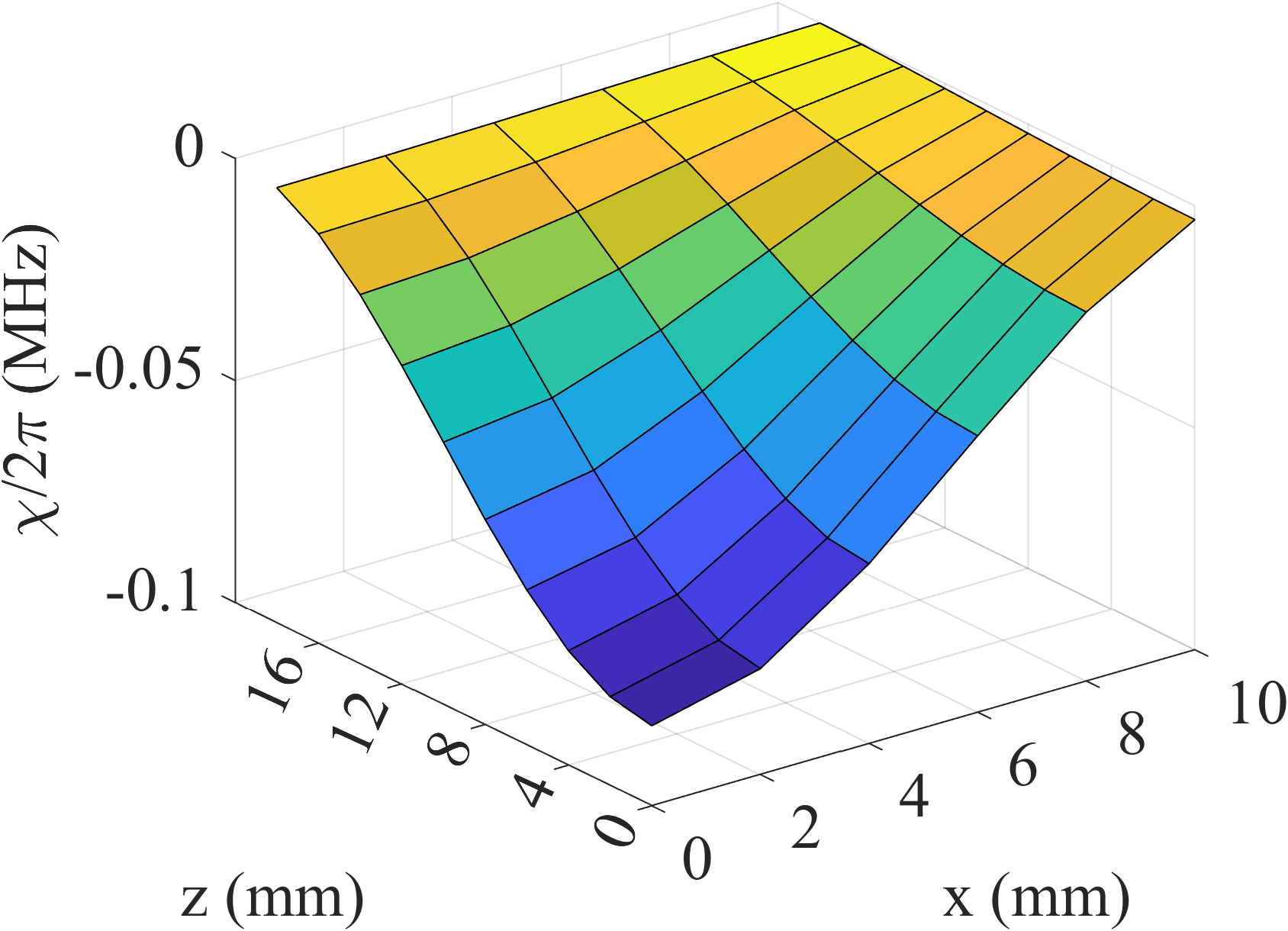}
		\caption{}
		\label{subfig:AC_stark_single_numerical}
    \end{subfigure}
    \begin{subfigure}[t]{0.49\linewidth}
		\includegraphics[width=\textwidth]{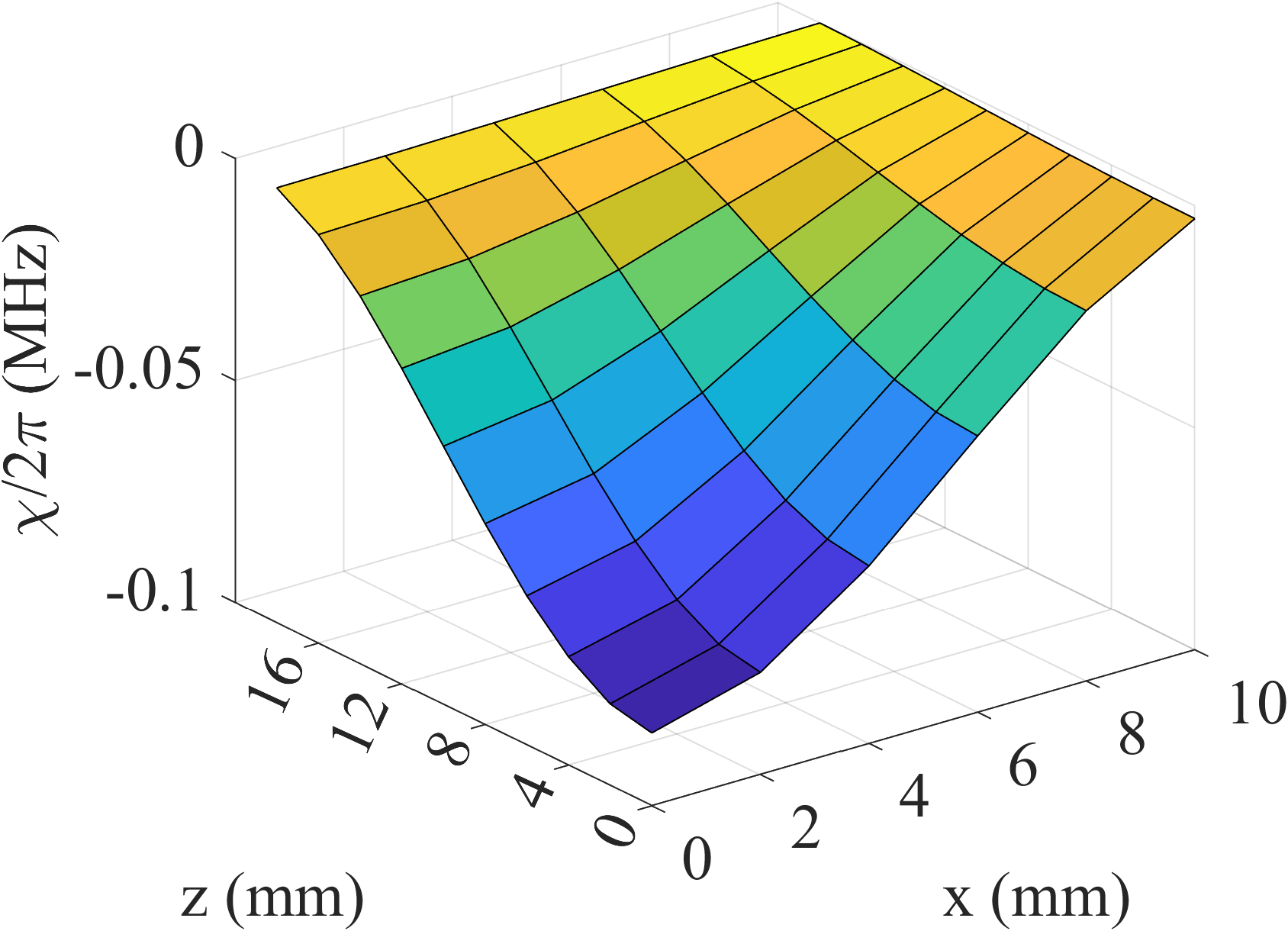}
		\caption{}
		\label{subfig:AC_stark_single_epr}
    \end{subfigure}
    \caption{AC-Stark shift $\chi$ computed while moving the transmon over one quadrant of the cross-sectional plane at a fixed vertical position in the cavity center for (a) analytical, (b) numerical eigenmode, and (c) EPR solution methods.}
    \label{fig:AC_Stark_shift_plots}
\end{figure}
 
Next, we show the effect of moving the transmon in the cavity on $\chi$ in Fig. \ref{fig:AC_Stark_shift_plots} when all methods use 8 Fock states per mode. Due to the similarity in frequencies, we compute the $\chi$ between the qubit and $\mathrm{TE}_{101}$ modes. Similar results are seen for all methods, highlighting that the approximations in our analytical solution do not break down so long as the dipole is kept a few millimeters from the cavity walls. We also summarize the relative error of all computed dispersive regime parameters in Table \ref{table:averaged-data}. Here, the system parameters are averaged as a function of the transmon location in the cavity. Relative errors are also computed between the methods as compared to the numerical EM eigenmode solution, which are shown in parentheses in Table \ref{table:averaged-data}. Given the typical experimental precision, the relative errors are all within reasonable ranges \cite{minev2021energy}. The main error source in our analytical solution is due to the dipole antenna capacitance computed from (\ref{eq:balanis dipole antenna}). With a better capacitance estimation, the analytical solution accuracy can be significantly improved, as will be shown shortly.

\begin{table}[t!]
    \renewcommand{\arraystretch}{1.5}
    \caption{Average system parameters and relative percent errors with respect to the numerical eigenmode data}
	\label{table:averaged-data}
    \centering
    \newcolumntype{D}{>{\centering\arraybackslash} m{0.075\textwidth}}
    \newcolumntype{M}{>{\centering\arraybackslash} m{0.10\textwidth}}
    \newcolumntype{P}{>{\centering\arraybackslash} m{0.10\textwidth}}
    \footnotesize
    \begin{tabular}{M|D|P|P}
    	\hline
    	\hline
    	\textbf{System Parameter} & \textbf{Numerical Eigenmode} & \textbf{EPR} &  \textbf{Analytical Solution} \\  
    	\hline
    	\hline
    	$\omega_{01}/2\pi$ (GHz) & 6.44 & 6.43 (0.21) & 6.39 (0.84)  \\
    	\hline
    	$\omega_1/2\pi$ (GHz) & 7.55 & 7.55 (8.82e-5) & 7.55 (1.5e-2)  \\
    	\hline
    	$\omega_2/2\pi$ (GHz) & 9.96 & 9.96 (8.88e-5) & 9.96 (2.6e-2)  \\
    	\hline
    	$\alpha/2\pi$ (MHz) & -379.00 & -360.78 (-4.81) & -371.72 (-1.92)  \\ 
        \hline
        $\chi/2\pi$ (MHz) & -0.025 & -0.026 (-2.23) & -0.028 (-10.38) \\
    	\hline	
    	\hline
     \end{tabular}
\end{table}

We now reconfigure the system to include two qubits that are located in the central plane of the cavity and are positioned underneath the coaxial probes. In these examples, we will compute the AC-Stark shift and $ZZ$-interaction rate as a function of one of the qubit frequencies by changing the value of $L_J$ while holding the other qubit's parameters fixed. Since our method does not explicitly include $L_J$'s in our EM eigenmode calculations, we only need to perform a single 3D FEM eigenmode analysis. In contrast to this, the EPR method includes $L_J$'s directly in the EM eigenmode calculations, and so requires running a new 3D FEM eigenmode analysis for every frequency point, which is inconvenient and computationally costly. Further, to achieve acceptable numerical convergence we needed to use seven Fock states for each mode in the EPR calculation, while our method only required three Fock states per mode. In both calculations, we use two qubit and three cavity modes. 

\begin{figure}[t!]
    \centering
    \includegraphics[width=0.85\linewidth]{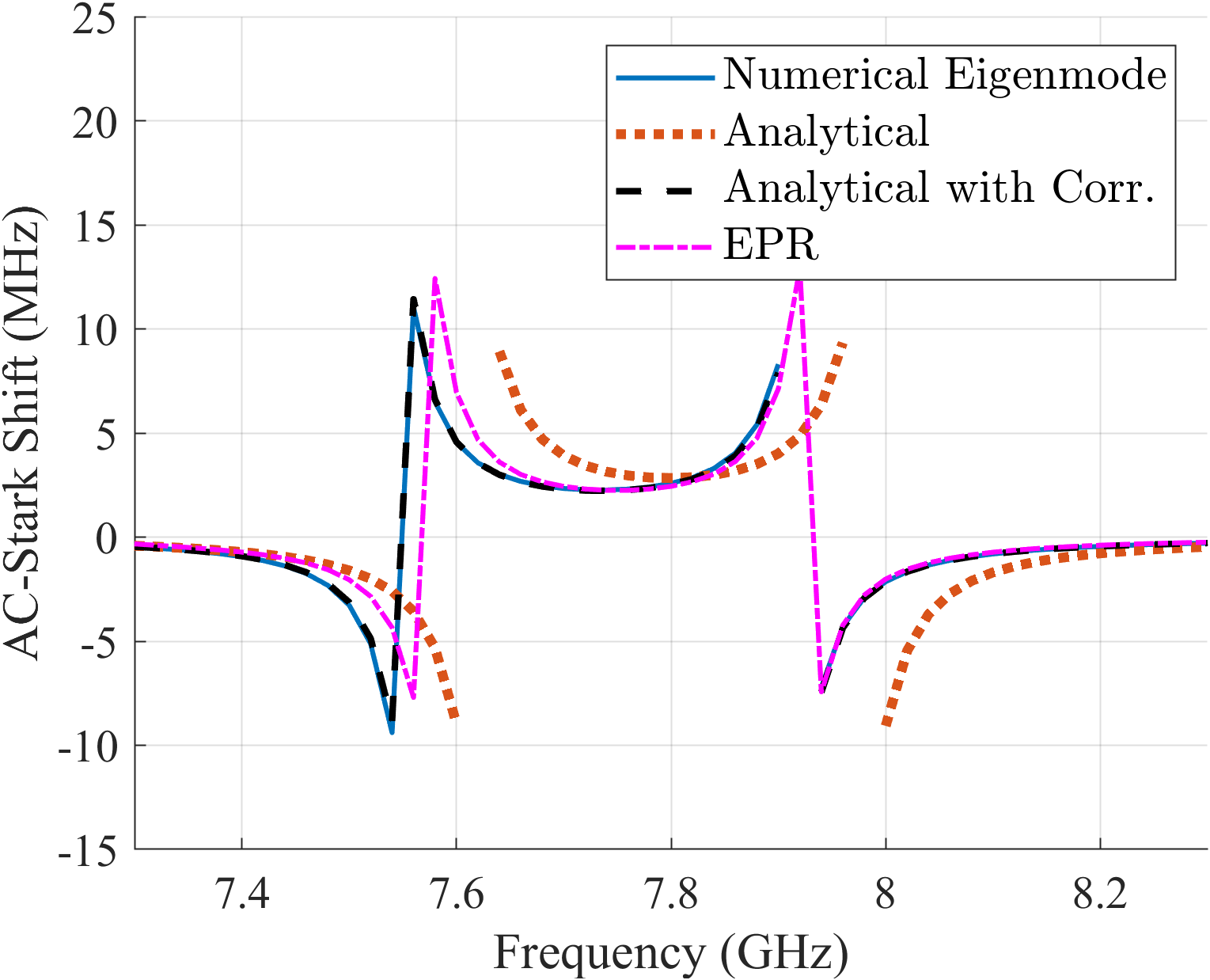}
    \caption{AC-Stark shift $\chi$ between one qubit and the $\mathrm{TE}_{101}$ mode as a function of qubit frequency using various methods, including an analytical calculation using a corrected dipole capacitance.}
    \label{fig:AC-stark-shift-multiqubit}
\end{figure}

We now sweep one qubit frequency from $7.3$ to $8.3 \, \mathrm{GHz}$ by varying it's $L_J$ from $7.420$ to $5.806 \, \mathrm{nH}$. We keep the second qubit fixed at a frequency of $6\, \mathrm{GHz}$ by setting it's $L_J$ to $10.756 \, \mathrm{nH}$. We then compute the AC-Stark shift between the qubit with varying frequency and the $\mathrm{TE}_{101}$ mode for each method discussed in this work. These are compared in Fig. \ref{fig:AC-stark-shift-multiqubit}, where it is seen that good agreement is obtained across the full frequency range. We also show that by improving the accuracy of the dipole antenna capacitance the error in our analytical solution can be greatly reduced. In particular, here we use a capacitance value computed via FEM for a dipole of the same characteristics located in free space (this modifies the capacitance from $9.091$ to $8.035 \, \mathrm{fF}$). In the results of Fig. \ref{fig:AC-stark-shift-multiqubit}, we also see ``resonant spikes'' in the AC-Stark shift that occur due to the qubit and cavity modes becoming resonant with one another. When this occurs, the dispersive regime approximations break down and the data can no longer be safely interpreted as an AC-Stark shift (gaps in the curves are due to this making identification of eigenvalues to calculate $\chi$ unreliable). However, nearby these resonant spikes do correspond to useful operating regimes, such as the ``straddling regime'' discussed in \cite{koch2007charge} that occurs between the two resonant spikes. 

For these simulations, all methods performed their analysis for $51$ frequency points. For our field-based approach using numerical eigenmode data, a single 3D FEM eigenmode calculation was required which took \textapprox$320$ seconds. This data was then used to generate $51$ Hamiltonian matrices whose eigenvalues and eigenvectors were then computed to evaluate the desired dispersive regime parameters, taking a total time of \textapprox$50$ seconds. Hence, the total simulation time for our method took \textapprox6 minutes. For the EPR method, $51$ 3D FEM eigenmode calculations were required which took \textapprox$320$ minutes and the corresponding $51$ Hamiltonian generation and eigenproblem solutions took \textapprox$1372$ minutes, leading to a total simulation time of over 28 hours. 

\begin{figure}[t!]
    \centering
    \includegraphics[width=0.85\linewidth]{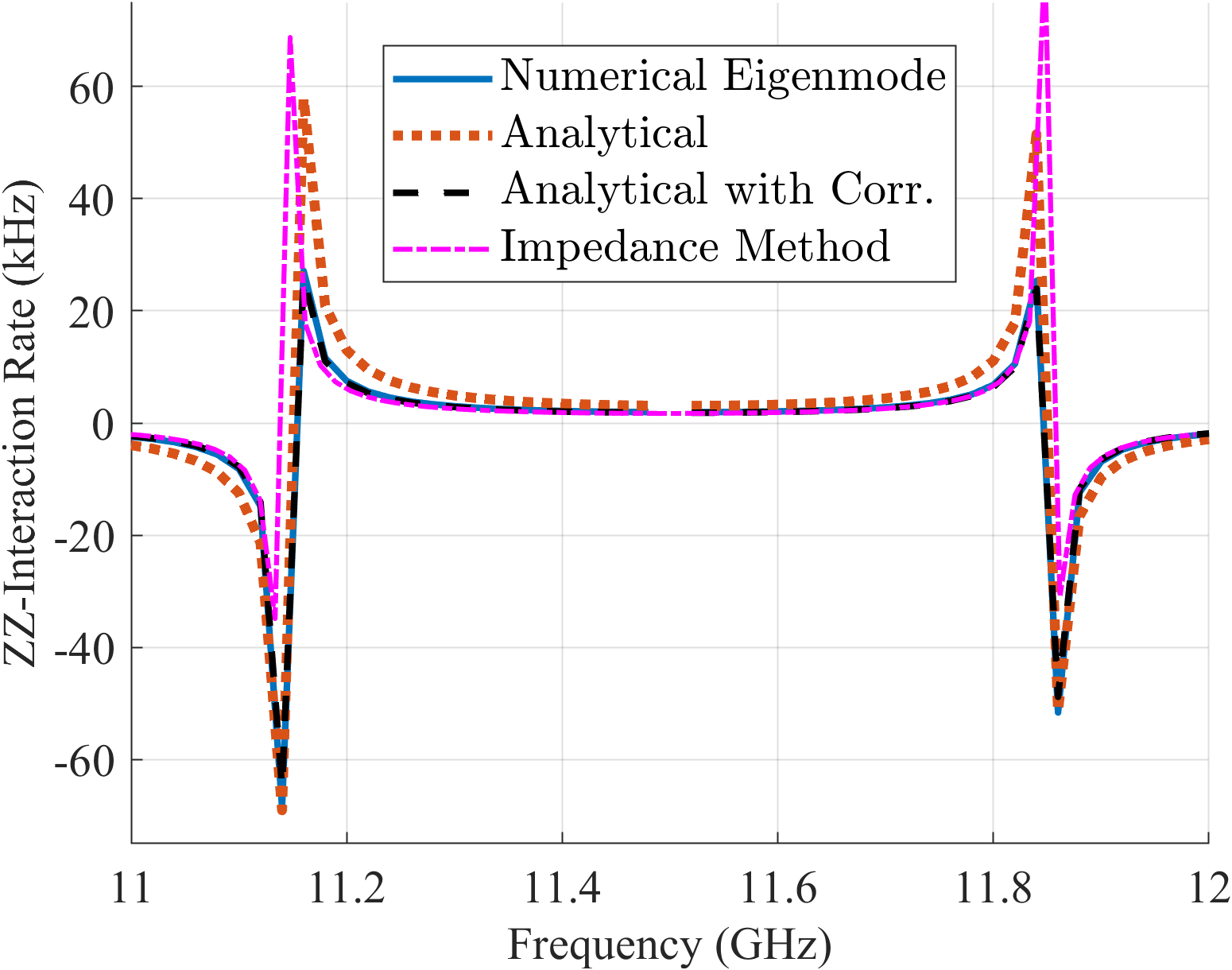}
    \caption{$ZZ$-interaction rate $\zeta$ between two qubits as a function of one qubit frequency computed using various methods, including an analytical calculation using a corrected dipole capacitance.}
    \label{fig:ZZ-interaction-multiqubit}
\end{figure}

Finally, we repeat a similar analysis but for computing the $ZZ$-interaction rate between the two qubits. To lead to a larger interaction strength, we fix one qubit frequency to $11.5 \, \mathrm{GHz}$ by setting $L_J$ to $3.095 \, \mathrm{nH}$ while varying the other qubit frequency from $11$ to $12 \,\mathrm{GHz}$ ($L_J$ ranging from $3.374$ to $2.850 \, \mathrm{nH}$). We set the frequencies in this way so that the $ZZ$-interaction can be facilitated primarily through the $\mathrm{TE}_{102}$ mode which has peaks at the locations of both qubits. To validate our analytical solution, we compare against our method using numerical EM eigenmodes and the impedance-based method of \cite{solgun2019simple,solgun2022direct}. From a theoretical analysis \cite{gambetta2013control}, we expect to see a similar shape to that seen in Fig. \ref{fig:AC-stark-shift-multiqubit} for the AC-Stark effect. As shown in Fig. \ref{fig:ZZ-interaction-multiqubit}, we do see this expected behavior and also see that all our calculation approaches agree well. Again, by improving the accuracy of the dipole antenna capacitance value we see that the accuracy of our analytical solution can be improved.

\section{Conclusion}
\label{sec:conclusion}
In this work, a 3D geometry was designed that could have all EM aspects needed in a quantum full-wave analysis evaluated using analytical techniques from classical EM theory. We first considered the analysis of an empty coaxial-fed rectangular waveguide cavity and showed how an analytical solution could be developed using quantum input-output theory. We then used this solution to calculate HOM interference curves, validating our approach by comparing it to a similar approach using numerical EM eigenmodes. Following this, we developed an analytical solution for a rectangular waveguide cavity with multiple transmon qubits embedded inside it. We validated our approach by computing key system parameters related to controlling qubits and compared them to our approach using numerical EM eigenmodes, as well as the independent EPR and impedance-based methods. This calculation also showed that our field-based formalism can be substantially more efficient than the EPR method in many commonly occurring situations. Future work can improve the accuracy of our analytical quantum full-wave solutions by using more sophisticated classical EM techniques and expanding our solution approach to other realistic geometries. Our field-based formalism can also be used to create a more complete numerical method to analyze complex dynamical effects in cQED systems.

\ifCLASSOPTIONcaptionsoff
  \newpage
\fi



%
\bibliographystyle{IEEEtran}
\bibliography{paper_main_bib}

%




\end{document}